\documentclass[epj, nopacs]{svjour}
\usepackage{latexsym}
\usepackage{amsmath}
\usepackage{graphics}
\usepackage{supertabular}
\begin{document}
\title{Generating QCD amplitudes in the color-flow basis with MadGraph}
\author{Kaoru Hagiwara\inst{1}\fnmsep\thanks{email:kaoru.hagiwara@kek.jp} \and Yoshitaro Takaesu\inst{1}\fnmsep\thanks{email:takaesu@post.kek.jp}
}                     

\institute{KEK Theory Center and Sokendai, Tsukuba, 305-0801, Japan}
\date{Received: date / Revised version: date}
%
\abstract{
We propose to make use of the off-shell
recursive relations with the color-flow decomposition in the calculation of QCD amplitudes on
MadGraph. We introduce colored quarks and their interactions with nine
gluons in the color-flow basis plus an Abelian gluon on MadGraph, such
that it generates helicity amplitudes in the color-flow basis with off-shell recursive
formulae for multi-gluon sub-amplitudes. We demonstrate calculations of up to 5-jet processes such as
$gg\rightarrow 5g$, $u\overline{u}\rightarrow 5g$ and $uu\rightarrow uuggg$. Although our
demonstration is limited, it paves the way to evaluate amplitudes with
more quark lines and gluons with Madgraph.
\PACS{
      {PACS-key}{discribing text of that key}   \and
      {PACS-key}{discribing text of that key}
     } 
} 
\maketitle
\section{Introduction}
\label{intro}
The Large Hadron Collider (LHC) has been steadily running at 7 TeV since the
end of March/2010. Since LHC is a hadron
collider, understanding of QCD radiation processes is essential for the success of the experiments. To evaluate new physics signals
and Standard Model (SM) backgrounds, we should resort to simulation programs which generate
events with many hadrons and investigate observable distributions. In each simulation,
one calculates matrix elements of hard processes with quarks and gluons,
and this task is usually done by an automatic amplitude calculator.

 MadGraph\cite{MG}
is one of those programs. Although it is highly developed and
has ample user-friendly utilities such as event generation with new physics signals
matched to Parton Shower\cite{MG/ME}, it cannot evaluate matrix
elements with more than five jets in the final state\footnote{In the case of
purely gluonic processes, even $gg\rightarrow 5g$ process cannot be
evaluated on MadGraph\cite{GPU2}.}\cite{GPU2}.

This is a serious drawback since exact evaluation of multi-jet matrix
elements is often required to study sub-structure of broad jets in
identifying new physics signatures over QCD background. It is also
disappointing because MadGraph generated HELAS amplitudes\cite{HELAS}
can be computed very fast on GPU (Graphic Processing Unit)\cite{GPU1,GPU2}.

Some of other simulation packages such as HELAC and Alpgen\cite{HELAC,Alpgen} employ QCD off-shell recursive
relations to produce multi-parton scattering
amplitudes efficiently. Successful computation of recursively evaluated
QCD amplitudes on GPU has been reported\cite{giele2010}. It is therefore interesting to examine the
possibility of implementing recursive relations for gluon currents in MadGraph without
sacrificing its benefits. In this paper, we examine the use of the
recursive amplitudes in the color-flow basis, since corresponding HELAS
amplitude package has been developed in ref. \cite{fabio}. The main purpose of
this paper is to show that this approach can be accommodated in MadGraph
with explicit examples.

The outline of this paper is as follows. In section~\ref{review}, we briefly
review the color-flow decomposition and off-shell recursive relations.
We then discuss the implementation of those techniques in
MadGraph, using its ``user mode'' option\cite{MG/ME} in section~\ref{implement}. In
section~\ref{total} and \ref{result}, we explain how we evaluate the color-summed amplitude
squared and show the numerical results of $n$-jet production cross section.
Section~\ref{conclusion} is devoted
to conclusions.

\section{The color-flow basis and off-shell recursive relations}
\label{review}
In this section we review the color-flow decomposition\cite{fabio} of QCD scattering amplitudes and the off-shell recursive relation\cite{berends} of gluonic currents in the color flow basis.
\subsection{The color-flow decomposition}
\label{colorflow}
First, we briefly review the color-flow decomposition. The Lagrangian of
QCD can be expressed as
\begin{equation}
 {\cal L}=-\frac{1}{4}({\cal F}^{\mu\nu})^j_i({\cal F}_{\mu\nu})^i_j
+\overline{\psi^j}\left\{i\gamma^{\mu}({\cal D}_{\mu})^i_j
-m\delta^i_j\right\}\psi_i,\label{lagrangian}
\end{equation}
where
\begin{align}
 ({\cal F}_{\mu\nu})^i_j=&\,\partial_{\mu}({\cal A}_{\nu})^i_j
-\partial_{\nu}({\cal A}_{\mu})^i_j \nonumber\\
&+i\frac{g}{\sqrt{2}}\left\{({\cal A}_{\mu})^k_j
({\cal A}_{\nu})^i_k-({\cal A}_{\nu})^k_j
({\cal A}_{\mu})^i_k\right\},
  \end{align}
\begin{equation}
 ({\cal
  D}_{\mu})^i_j=\partial_{\mu}\delta^i_j+i\frac{g}{\sqrt{2}}({\cal
  A}_{\mu})^i_j,
\end{equation}
\begin{equation}
 ({\cal A}_{\mu})^i_j=\sqrt{2}A_{\mu}^a(T^a)^i_j.
\end{equation}
\begin{equation}
 (T^a)^i_j(T^b)^j_i = \frac{1}{2}\delta^{ab}.
\end{equation}
Upper and lower indices are those of ${\bf \overline{3}}$ and ${\bf 3}$
representations, respectively. Note that the gluon fields, $({\cal
A}_{\mu})^i_j\,\,(i,j=1, 2, 3)$, are renormalized by a
factor of $\sqrt{2}$, and hence the coupling $g$ is divided by
$\sqrt{2}$. At this stage, not all the nine gluon fields are independent
because of  the traceslessness of the $SU(3)$ generators,
\begin{equation}
({\cal A}_{\mu})^1_1 + ({\cal A}_{\mu})^2_2 + ({\cal A}_{\mu})^3_3 = 0.
\end{equation}

Here we introduce the ``Abelian gluon'',
\begin{align}
 ({\cal B}_{\mu})^i_j=\sqrt{2}B_{\mu}\frac{\delta^i_j}{\sqrt{2N}}\,\,,
\end{align}
This is essentially the gauge boson , $B_{\mu}$, of $U(1)$ subgroup of $U(3)$, combined with its generator, $\delta^i_j/\sqrt{2N}$, which is normalized to 1/2.  We then rewrite
eq.~(\ref{lagrangian}) by adding and subtracting the Abelian gluon,
such that the QCD Lagrangian is expressed as
\begin{align}
 {\cal L}=&-\frac{1}{4}({\cal F}^{\mu\nu})^j_i({\cal
 F}_{\mu\nu})^i_j
-\frac{1}{4}({\cal B}^{\mu\nu})^j_i({\cal
 B}_{\mu\nu})^i_j
+\frac{1}{4}({\cal B}^{\mu\nu})^j_i({\cal B}_{\mu\nu})^i_j\nonumber\\
&+\overline{\psi^j}\Biggl[\gamma^{\mu}\left\{i\partial_{\mu}\delta^i_j-\frac{g}
{\sqrt{2}}({\cal A}_{\mu})^i_j-\frac{g}{\sqrt{2}}({\cal
  B}_{\mu})^i_j+\frac{g}{\sqrt{2}}({\cal
  B}_{\mu})^i_j\right\}\nonumber \\
&-m\delta^i_j\Biggr]\psi_i\nonumber\\
=&-\frac{1}{4}({\cal G}^{\mu\nu})^j_i({\cal
 G}_{\mu\nu})^i_j
-\left\{-\frac{1}{4}({\cal B}^{\mu\nu})^j_i({\cal
 B}_{\mu\nu})^i_j\right\}\nonumber\\
&+\overline{\psi^j}\Biggl[\gamma^{\mu}\left\{i\partial_{\mu}\delta^i_j-\frac{g}
{\sqrt{2}}({\cal
  G}_{\mu})^i_j+\frac{g}{\sqrt{2}}({\cal
  B}_{\mu})^i_j\right\}-m\delta^i_j\Biggr]\psi_i
\label{eq:cflag},
\end{align}
with
\begin{align}
({\cal G}_{\mu})^i_j=({\cal A}_{\mu})^i_j+({\cal B}_{\mu})^i_j.
\end{align}
Here all the nine gluons, $({\cal G}_{\mu})^i_j,$ are now independent while the
 covariant tensors keep the same form

\begin{align}
({\cal G}_{\mu\nu})^i_j=&\partial_{\mu}({\cal G}_{\nu})^i_j
-\partial_{\nu}({\cal G}_{\mu})^i_j \nonumber\\
&+i\frac{g}{\sqrt{2}}\left\{({\cal G}_{\mu})^k_j
({\cal G}_{\nu})^i_k-({\cal G}_{\nu})^k_j
({\cal G}_{\mu})^i_k\right\}\label{eq:gmunu},\\
({\cal B}_{\mu\nu})^i_j=&\partial_{\mu}({\cal B}_{\nu})^i_j
-\partial_{\nu}({\cal B}_{\mu})^i_j\label{eq:bmunu}.
\end{align}
 In this basis, these nine gluons have one-to-one correspondence to a set
of indices of ${\bf 3}$
and ${\bf \overline{3}}$ representations of $U(3)$, $(j,i) \iff ({\cal G}_{\mu})^i_j$, and we address them and this basis as $U(3)$ gluons  and the color-flow basis, respectively, in the following discussions. Although the definition of the covariant tensor (\ref{eq:gmunu}) contains self-coupling
 terms for $U(3)$ gluons, the Abelian gluon contribution actually drops out
 in the sum. Accordingly, the Feynman rules for $U(3)$ gluons give directly the $SU(3)$ (QCD) amplitudes for purely
 gluonic processes\cite{fabio}. 

\begin{figure}
\centering
\resizebox{0.49\textwidth}{!}{
\includegraphics{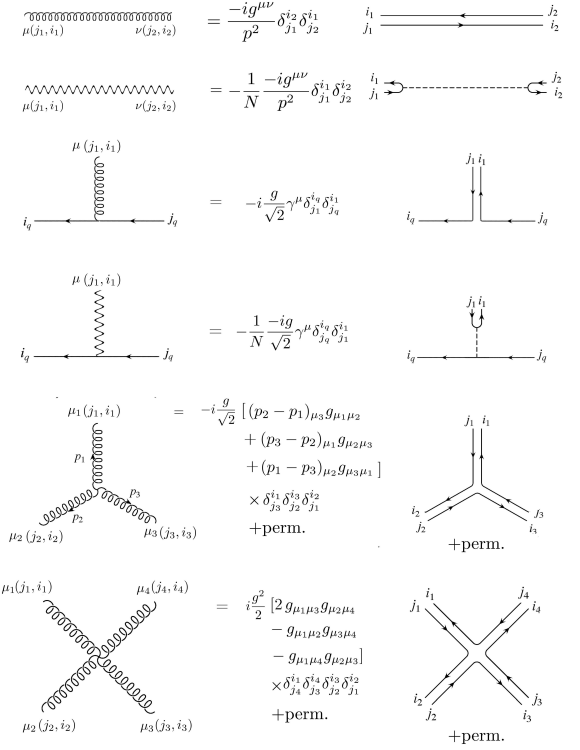}
}
\caption{QCD Feynman rules in the color-flow basis. The standard
Feynman diagrams are shown in the left and the color-flow diagrams are
 shown in the right. The wavy lines in the left and the dashed lines in the right denote the propagating Abelian gluon. The
 gluonic three point vertex has an additional non-cyclic
 permutation in color indices, while the four point vertex has additional five non-cyclic permutations.}
\label{fig:diagrams}
\end{figure}

We list the Feynman rules derived from the Lagrangian  (\ref{eq:cflag}) in
Fig.\,\ref{fig:diagrams}. 
It should be noted that both the propagator and the quark vertex of the Abelian gluon have an extra
$-1/N$ factor because of the color-singlet projector, $\delta^i_j/\sqrt{N}$. The negative sign of this factor directly comes from the sign
of the Abelian gluon kinetic term in the Lagrangian. We can evaluate QCD amplitudes by using these Feynman rules.
 
\begin{figure}
\resizebox{0.49\textwidth}{!}{%
\includegraphics{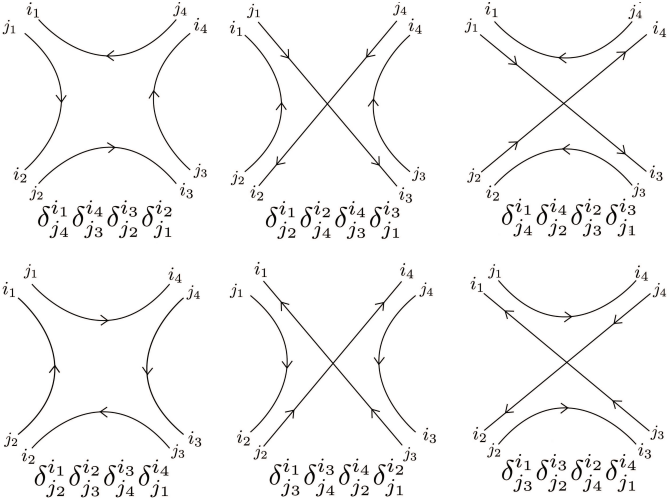}
}
\caption{All 6 color flows of the process ${\cal
G}_{j_1}^{i_1}{\cal G}_{j_2}^{i_2}\rightarrow{\cal G}_{j_3}^{i_3}{\cal
G}_{j_4}^{i_4}$.}
\label{fig:colorflows}
\end{figure}

 For a $n$-gluon scattering process, it is written as
\begin{align}
 {\cal M}(ng)=&\sum_{\sigma\in S_n/Z_n}\delta^{i_{\sigma(n)}}_{j_{\sigma(n-1)}}
\delta^{i_{\sigma(n-1)}}_{j_{\sigma(n-2)}}\cdots\delta^{i_{\sigma(1)}}_{j_{\sigma(n)}}\nonumber\\
&\hspace{4em}\times A_{\sigma}\Bigl(\sigma(1),\sigma(2),\cdots,\sigma(n)\Bigr),
\label{eq:Mn}
\end{align}
obtained from the corresponding $n$ $U(3)$ gluon scattering process. $A_{\sigma}\Bigl(\sigma(1),\sigma(2),\cdots,\sigma(n)\Bigr)$ are
gauge-invariant partial
amplitudes, which depend only on particle momenta and
helicities represented simply by the gluon indices, $1,\cdots,n$, and
 the summation is taken over all $(n-1)!$ non-cyclic permutations of
gluons. A set of $n$ Kronecker delta's in each term gives the "color flow"
of the corresponding partial amplitudes. As an example, we list in Fig.\,\ref{fig:colorflows} all six color flows indicated by Kronecker delta's
for the  $gg\rightarrow gg$ process, ${\cal
G}_{j_1}^{i_1}{\cal G}_{j_2}^{i_2}\rightarrow{\cal G}_{j_3}^{i_3}{\cal
G}_{j_4}^{i_4}$.

\begin{figure}
\resizebox{0.49\textwidth}{!}{%
\includegraphics{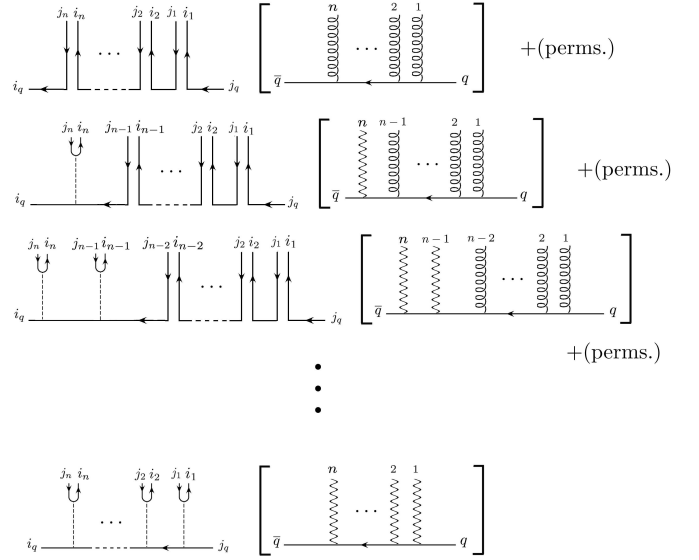}
}
\caption{Color flows of $q\overline{q}\rightarrow ng$ process. For each
 color flow, one of the corresponding Feynman diagrams is shown in the parenthesis. The
 term ``perms.'' stands for the other color flows obtained by permutations among gluons.}
\label{fig:qqbarcf}
\end{figure}

On the other hand, if a scattering process involves quarks, contributions
from the Abelian gluon do not decouple. Therefore, we have to add all
its contributions to the total amplitude. For
processes with one quark line, such contributions come from Abelian gluon
emitting diagrams. The total
amplitude for $q\overline{q}\rightarrow ng$ process becomes
\begin{align}
{\cal M}(q&\overline{q}+ng)\nonumber\\
=&\sum_{\sigma\in S_n}\delta^{i_q}_{j_{\sigma(n)}}
\delta^{i_{\sigma(n)}}_{j_{\sigma(n-1)}}\cdots\delta^{i_{\sigma(1)}}_{j_q} A^0_{\sigma}\Bigl(q,\sigma(1),\cdots,\sigma(n),\overline{q}\Bigr)\nonumber\\
&+\sum_{\sigma\in S_n}\left(-\frac{1}{N}\right)\delta^{i_{\sigma(n)}}_{j_{\sigma(n)}}\delta^{i_q}_{j_{\sigma(n-1)}}
\cdots\delta^{i_{\sigma(1)}}_{j_q}\nonumber\\
&\hspace{6em}\times A^1_{\sigma}\Bigl(q,\sigma(1),\cdots,\sigma(n-1),\overline{q};\,\sigma(n)\Bigr)\nonumber\\
&+\frac{1}{2!}\sum_{\sigma\in
 S_n}\left(-\frac{1}{N}\right)^2\delta^{i_{\sigma(n)}}_{j_{\sigma(n)}}\delta^{i_{\sigma(n-1)}}_{j_{\sigma(n-1)}}
 \delta^{i_q}_{j_{\sigma(n-2)}}\cdots\delta^{i_{\sigma(1)}}_{j_q}\nonumber\\ 
&\hspace{2em}\times A^2_{\sigma}\Bigl(q,\sigma(1),\cdots,\sigma(n-2),\overline{q};\,\sigma(n-1),\sigma(n)\Bigr)\nonumber\\
&\hspace{0.5em}\vdots\nonumber\\
&+\left(-\frac{1}{N}\right)^n\delta^{i_n}_{j_n}\cdots\delta^{i_1}_{j_1}\delta^{i_q}_{j_q}
A^n_{\sigma}\left(q,\overline{q};\,1,\cdots,n\right),
\label{eq:qqng}
\end{align}
where
$A^k_{\sigma}\Bigl(q,\sigma(1),\cdots,\sigma(n-k),\overline{q};\,\sigma(n-k+1),\cdots,\sigma(n)\Bigr)$
is the partial amplitude with $k$ Abelian gluons, and $q$ and
$\overline{q}$ denote the momenta and the helicities of the quark and the anti-quark,
respectively. The summation is taken over all the $n!$ permutations of $n$ gluons. The
first term in the r.h.s. of eq.~(\ref{eq:qqng}) gives contributions from
$n$ $U(3)$ gluons, and the other terms give contributions from
$(n-k)$ $U(3)$ gluons and $k$ Abelian gluons, summed over $k=1$ to
$n$, where $(-1/N)^k$ comes from the $-1/N$ factor in the Abelian gluon vertex as depicted in Fig.~\ref{fig:diagrams}. Corresponding color flows of those partial amplitudes are shown in Fig.\,\ref{fig:qqbarcf}.

\begin{figure}
\resizebox{0.49\textwidth}{!}{%
\includegraphics{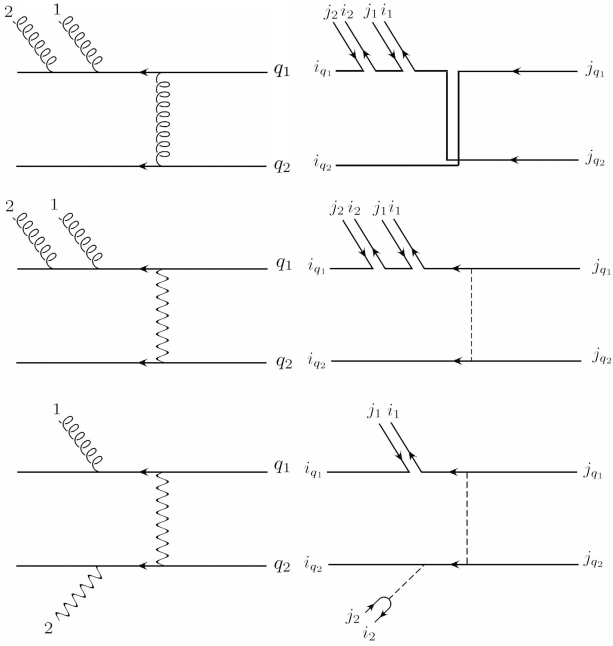}
}
\caption{Sample diagrams for the $q_1q_2\rightarrow q_1q_2gg$ process. Each
 Feynman diagram in left-hand-side is accompanied by the corresponding
 color-flow diagram in the right-hand-side.}
\label{fig:2qqdiagrams}
\end{figure}

For processes with two or more quark lines, the Abelian gluon can be
exchanged between them, giving the amplitudes with at
least $-1/N$
suppression.  Therefore, the total amplitude of processes with two quark lines such as  $q_1q_2\rightarrow q_1q_2+n$ gluons becomes
\begin{align}
{\cal M}&(q_1\overline{q_1}+q_2\overline{q_2}+ng)\nonumber\\
&=\sum_{\sigma\in S_n}\sum_{r=0}^n\nonumber\\
&\left[\left(\delta^{i_{q_1}}_{j_{\sigma(n)}}
\delta^{i_{\sigma(n)}}_{j_{\sigma(n-1)}}\cdots\delta^{i_{\sigma(r+1)}}_{j_{q_2}}\right)\left(\delta^{i_{q_2}}_{j_{\sigma(r)}}
\delta^{i_{\sigma(r)}}_{j_{\sigma(r-1)}}\cdots\delta^{i_{\sigma(1)}}_{j_{q_1}}\right)\right.\nonumber\\
&\times A^0_{\sigma}\Bigl(q_1,\sigma(1),\cdots,\sigma(r),\overline{q_2}\,|\,q_2,\sigma(r+1),\cdots,\sigma(n),\overline{q_1}
\Bigr)\nonumber\\
\vspace{2em}
&-\frac{1}{N}\left(\delta^{i_{q_2}}_{j_{\sigma(n)}}
\delta^{i_{\sigma(n)}}_{j_{\sigma(n-1)}}\cdots\delta^{i_{\sigma(r+1)}}_{j_{q_2}}\right)
\left(\delta^{i_{q_1}}_{j_{\sigma(r)}}
\delta^{i_{\sigma(r)}}_{j_{\sigma(r-1)}}\cdots\delta^{i_{\sigma(1)}}_{j_{q_1}}\right)
\nonumber\\
&\Bigl.\times B^0_{\sigma}\Bigl(q_1,\sigma(1),\cdots,\sigma(r),\overline{q_1}\,|\,q_2,\sigma(r+1),\cdots,\sigma(n),\overline{q_2}
\Bigr)\Bigr]\nonumber\\
&+\left({\rm partial\,\, amplitudes\,\, with \,\,external\,\,
 Abelian\,\, gluons}\right).
\label{eq:2qqng}
\end{align}
$A^0_{\sigma}\Bigl(q_1,\sigma(1),\cdots,\sigma(r),\overline{q_2}\,|\,q_2,\sigma(r+1),\cdots,\sigma(n),\overline{q_1}\Bigr)$
(or $B^0_{\sigma}$) denotes a partial
amplitude with $n$ $U(3)$ gluons, where two sets of arguments divided by ``$|$'' belong to
two different color-flow chains; one starts from $q_1$ ($q_1$ in the initial
state) and ends with $\overline{q_2}$ ($q_2$ in the final state), and the
other starts from $q_2$ and ends with $\overline{q_1}$, which are given
explicitly by two
sets of delta's before the partial amplitude. The difference between
$A^0_{\sigma}$ and $B^0_{\sigma}$ is that the former consists of $U(3)$-gluon-exchange diagrams
while the latter consists of Abelian-gluon-exchange ones.
Here we write down explicitly the partial amplitudes with external
$U(3)$ gluons,
$A^0$ and $B^0$, whereas those with external Abelian gluons should be
added as in the previous one-quark-line case. For illustration, some typical
diagrams for the $q_1q_2\rightarrow q_1q_2gg$ process
are shown in Fig.\,\ref{fig:2qqdiagrams}.

\subsection{Recursive relations of off-shell gluon currents}
\label{recursive}
\begin{figure}[b]
\resizebox{0.49\textwidth}{!}{%
\includegraphics{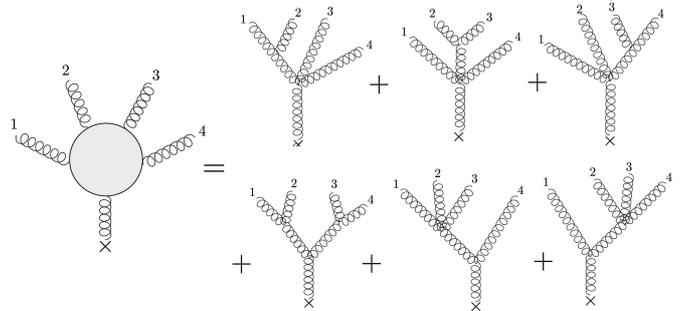}
}
\caption{The 5-point gluon off-shell current. Numbers denote gluon momenta
 and helicities, and the curly line with $\times$ symbol denotes the off-shell
 gluon.}
\label{fig:5offshell}
\end{figure}

Next, we mention the off-shell recursive relations in the color-flow
basis. We define a $(n+1)$\,-point off-shell gluon current $J^{\mu}(1, 2,\cdots,n;\,x)$ recursively
as
\begin{equation}
J^{\mu}(k;\,x)\equiv\delta_{j_k}^{\,i_x}\delta_{j_x}^{\,i_k}\epsilon^{\mu}(k)\hspace{1em}(1\leq
 k\leq n),\label{j2}
 \end{equation}
 \begin{align*}
J^{\mu}&(l, l+1,\cdots,m;\,x) \hspace{2em}(1\leq l < m\leq n)\nonumber\\
&\equiv\delta^{\,i_x}_{j_m}\delta^{\,i_m}_{j_{m-1}}\cdots\delta^{\,i_{l}}_{j_x}J^{\mu}(l, l+1,\cdots,m) \nonumber\\
&\equiv\frac{-i}{P^2_{l, m}}\Biggl\{\sum_{y, z}\sum_{k=l}^{m-1}
\delta_{\,i_z}^{\,i_x}\delta_{\,i_y}^{j_z}\delta_{j_x}^{j_y}\,V^{\nu\rho\mu}_3(P_{l, k},P_{k+1, m},-P_{l, m})\nonumber\\
 &\hspace{7em}\times J_{\nu}
(l,\cdots,k;\,y)J_{\rho}(k+1,\cdots,m;\,z)\Biggr.\nonumber\\
&\hspace{0.5em}+\sum_{y, z, w}\sum_{k=l}^{m-2}\sum_{q=k+1}^{m-1}
\delta_{\,i_w}^{\,i_x}\delta_{\,i_z}^{j_w}\delta_{\,i_y}^{j_z}\delta_{j_x}^{j_y}
 \,V^{\nu\rho\sigma\mu}_4J_{\nu}(l, \cdots,k;\,y)\nonumber
\end{align*}
\begin{align}
 &\hspace{2em}\Biggl.\times J_{\rho}(k+1,\cdots,q;\,z)J_{\sigma}(q+1,\cdots,m;\,w)\Biggr\},
\label{eq:off-shell}
\end{align}
where the numbers in the argument of currents represent gluons,
 and their order respects the color flow as in the partial amplitudes,
 $A_{\sigma}$, in eq.~(\ref{eq:Mn}); the argument after semicolon denotes the off-shell gluon. As an initial value, the 2-point gluon current,
 $J^{\mu}(k;\,x)$, is defined as the polarization vector of the gluon $k$,
 $\epsilon^{\mu}(k)$. $P_{i,j} (i<j)$ is the partial sum
 of gluon momenta:
 \begin{equation}
P_{i,j}=p_i+p_{i+1}+\cdots+p_j,
 \end{equation}
where $p_i$ to $p_j$ are defined as
 flowing-out momenta.
$V_3^{\nu\rho\mu}$ and $V_4^{\nu\rho\sigma\mu}$ denote the three- and
 four-point gluon vertices from the Feynman rule of Fig.\,\ref{fig:diagrams}:
\begin{align}
V_3^{\nu\rho\mu}(p,q,r)=&-i\frac{g}{\sqrt{2}}\bigl\{(q-p)^{\mu}g^{\nu\rho}+(r-q)^{\nu}g^{\rho\mu}\nonumber\\
&\hspace{8.7em}+(p-r)^{\rho}g^{\mu\nu}\bigr\},\\
V_4^{\nu\rho\sigma\mu}=&\,\,i\frac{g^2}{2}(2g^{\nu\sigma}g^{\rho\mu}-g^{\nu\rho}g^{\sigma\mu}-g^{\nu\mu}g^{\rho\sigma}).
\end{align}
The summation over $y,z$ and $w$ in eq.~(\ref{eq:off-shell}) are taken over the {\bf 3} and $\overline{{\bf 3}}$ indices, $(j_y, i_y), (j_z, i_z)$
 and $(j_w, i_w)$, of off-shell gluons in currents $J_{\nu}, J_{\rho}$ and $J_{\sigma}$, respectively.
 For
illustration, the 5-point off-shell current is shown explicitly in
Fig.\,\ref{fig:5offshell}.

Helicity amplitudes of pure gluon processes in the color-flow basis are obtained from off-shell
gluon currents as
\begin{align}
\delta&_{j_{n-1}}^{i_n}\delta_{j_{n-2}}^{i_{n-1}}\cdots\delta_{j_n}^{i_1}\,A(1,\cdots,n)\nonumber\\
&=\left\{\sum_{x,y,z}\sum_{k=1}^{n-2}\sum_{l=k+1}^{n-1}
\delta_{i_z}^{j_x}\delta_{i_y}^{j_z}\delta_{i_x}^{j_y}\,V^{\mu\nu\rho}_3(P_{1,k}, P_{k+1,l},P_{l+1,n})\right.\nonumber\\
&\times J_{\mu}(1,\cdots,k;\,x)J_{\nu}(k+1,\cdots,l; y) J_{\rho}(l+1,\cdots,n;\,z)\nonumber\\
&+\sum_{x,y,z,w}\sum_{k=1}^{n-3}\sum_{l=k+1}^{n-2}\sum_{m=l+1}^{n-1}
\delta_{i_w}^{j_x}\delta_{i_z}^{j_w}\delta_{i_y}^{j_z}\delta_{i_x}^{j_y}\,
V^{\mu\nu\rho\sigma}_4\nonumber\\
& \times J_{\mu}(1,\cdots,k;x) J_{\nu}(k+1,\cdots,l;y) J_{\rho}(l+1,\cdots,m;z) \nonumber\\
&\hspace{11em}\Biggl. \times J_{\sigma}(m+1,\cdots,n;w)\Biggr\}.
\label{eq:amplitude}
\end{align}
Starting from the external wave functions (\ref{j2}), the recursive
relation computes partial amplitudes very effectively by using the
HELAS code. When the off-shell line ``$\times$'' in Fig.\,\ref{fig:5offshell} is set
on-shell, the same set of diagrams give the complete helicity amplitudes
in the color-flow basis.

\section{{\large Implementation in MadGraph}}
\label{implement}

In this section, we discuss how we implement off-shell recursive
relations in MadGraph in the color-flow
basis and how we generate helicity amplitudes of QCD processes.

\subsection{Subroutines for off-shell recursive formulae}
\label{subroutine}
\begin{figure}
\centering
\resizebox{0.4\textwidth}{!}{%
\includegraphics{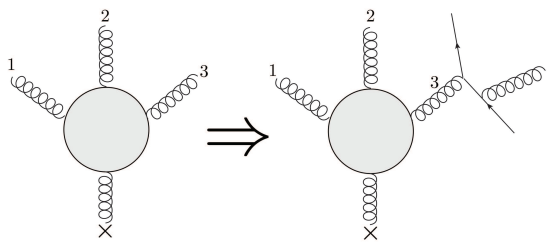}
}
\caption{An example of the 4-point off-shell
 current (the external leg with the cross symbol is off-shell), where an
 on-shell gluon in the current (labeled as 3) is replaced by an off-shell gluon attached to
 a quark line.}
\label{fig:replace}
\end{figure}

First, we introduce new HELAS\cite{HELAS} subroutines in MadGraph which make
  $n$-point gluon off-shell currents and $n$-gluon amplitudes in the color-flow basis, according to
 eq.~(\ref{eq:off-shell}) and eq.~(\ref{eq:amplitude}),
  respectively. Although the expression eq.~(\ref{eq:off-shell}) gives
  the off-shell gluon current made from $n$
  on-shell gluons, in HELAS amplitudes any input on-shell gluon wave
  functions can be replaced by arbitrary off-shell gluon currents with the same
  color quantum numbers. Shown in
  Fig.\,\ref{fig:replace} is an
  example of such diagrams that are calculated by the HELAS subroutine for the
  4-point off-shell gluon current.
Thanks to this property of the HELAS
  amplitudes, we need to introduce only two types of subroutines: one which computes
   helicity amplitudes of $n$-gluon processes via
  eq.~(\ref{eq:amplitude}) and the other which computes off-shell gluon
  currents from $n$-point gluon vertices via eq.~(\ref{eq:off-shell}). We
  name the first type of subroutines as {\tt gluon\#} and the second
  type as
  {\tt jgluo\#}, where {\tt \#} denotes the number of external on-shell
  and off-shell gluons.

The number of new subroutines we should
  add to MadGraph depends on the number of external partons (quarks and
  gluons) in QCD
  processes. Processes with $n$-partons can be classified as those
  with $n$ gluons, those with $(n-2)$ gluons and one quark line, those
  with $(n-4)$ gluons and two quark lines, and so on.

In the color-flow basis,
  the first class of processes with $n$ external gluons are calculated
  by just one amplitude subroutine, {\tt gluon\,n}.

For the
  second class of processes with $(n-2)$ gluons and one quark line, we
  need up to $(n-1)$-point off-shell current subroutines, {\tt jgluo\#}
  with {\tt \#} $=4$ to $n-1$. This is because the largest off-shell
  gluon current appears in the computation of diagrams where $(n-2)$ on-shell gluons are connected to
  the quark line through one off-shell gluon, which can be computed by the $(n-1)$-point off-shell
  current and the $q\overline{q}g$ amplitude subroutine, {\tt
  iovxxx}. Note that the same diagram can be computed by the off-shell
  gluon current subroutine made by the quark pair, {\tt jioxxx}, and the
  $(n-1)$-point gluon amplitude subroutine, {\tt gluon\,(n-1)}. This
  type of redundancy in the computation of each Feynman diagram is
  inherent in the HELAS amplitudes, which is often useful in testing the
  codes.

 For $n$-parton processes with $(n-4)$ gluons and two quark lines, we need
up to $(n-2)$-point off-shell current subroutines. By also introducing
multiple gluon amplitude subroutines up to $(n-2)$-point vertex, the maximum
redundancy of HELAS amplitudes is achieved. Likewise, for $n$-parton
processes with $m$ quark lines, we need up to $(n-m)$-point off-shell
current or amplitude subroutines.
 
 We list in Table \ref{tb:subroutines} the new HELAS subroutines we
 introduce in this study.
\begin{table}
\begin{tabular}{r|l}
\hline\hline
{\tt gluon\#}&{\tt \#}-gluon amplitude in the color-flow basis\\
{\tt jgluo\#}&off-shell gluon current from ({\tt \#}$-1)$ external glu-\\
& ons in the color-flow basis\\\\
{\tt ggggcf}&4-gluon amplitude from the contact 4-gluon ver-\\
&tex in the color-flow basis\\
{\tt jgggcf}&off-shell gluon current from the contact 4-gluon\\
& vertex in the color-flow basis\\\\
 {\tt jioaxx}&off-shell Abelian gluon current from a quark pair\\
 & in the color-flow basis\\
\hline\hline
\end{tabular}\\
\hspace{2em}
\caption{New HELAS subroutines added into
 MadGraph. {\tt gluon\#} and {\tt jgluo\#} compute the {\tt \#}-gluon
 amplitude and the off-shell gluon
 current from the \mbox{{\tt \#}-gluon} vertex, respectively, by using the off-shell
 recursive formulae in the color-flow basis. We use {\tt gluon\#} with
 {\tt \#} $=4$
 to $7$ and {\tt jgluo\#} with {\tt \#} $=4$ to $6$ in this study. Two subroutines for the contact 4-gluon vertex ({\tt ggggcf} and {\tt
 jgggcf}) are introduced to sum over the two channels ($s$ and $t$ or
 $s$ and $u$) for a given color flow, according to the Feynman rule in
 Fig.\,\ref{fig:diagrams}.  We also add the off-shell Abelian
 gluon current subroutine from a quark pair ({\tt jioaxx}).}
\label{tb:subroutines}
\end{table}
The subroutine {\tt gluon\#} evaluates a {\tt \#}-gluon
  amplitude in the color-flow basis, and {\tt jgluo\#} computes an off-shell current from
  ({\tt \#}$-1$) external gluons. Since we consider up to seven parton processes, $gg\rightarrow 5g$, $u\overline{u}\rightarrow
  5g$ and $uu\rightarrow uu3g$, in this study, we use {\tt
  gluon4} to {\tt gluon7} and {\tt jgluo4} to {\tt jgluo6}.

In addition to these subroutines that computes amplitudes and currents
recursively, we also introduce three subroutines: {\tt ggggcf}, {\tt
jgggcf} and {\tt jioaxx}. Two of them, {\tt ggggcf} and {\tt jgggcf}, evaluate  an
amplitude and an off-shell current from the contact 4-gluon vertex,
following the Feynman rule of Fig.\,\ref{fig:diagrams}. Although
 the amplitude subroutine and the off-shell current subroutine for the 4-gluon vertex already exist in MadGraph, {\tt
ggggxx} and {\tt jgggxx}, we should introduce
the new ones which evaluate the sum of s- and t-type or s- and u-type
vertices
 for a given color flow, since the
default subroutines compute only one type at a time. {\tt jioaxx}
computes an off-shell Abelian gluon current made by a quark pair and is
essentially the same as the off-shell gluon current
 subroutine, {\tt jioxxx}, in the HELAS library\cite{HELAS} except for an extra $-N$ factor:
 \begin{equation}
  {\tt jioaxx} = -N \times {\tt jioxxx}.
\end{equation}
Note that introducing this $-N(=-1/N \times N^2)$ factor is equivalent to summing up
contributions from all  Abelian gluon propagators, as we discussed in section~\ref{review}. We show the codes
of ggggcf, jgggcf, gluon5 and jgluo5 in Appendix.

\subsection{Introduction of a new Model: CFQCD}
\label{cfqcd}
Next, we define a new Model with which MadGraph generates HELAS
amplitudes in the color-flow basis since the present MadGraph computes them in the color-ordered basis\cite{multiparton}.
 A Model is a
set of definitions of particles,
their interactions and couplings; there are about twenty preset Models in
the present MadGraph package\cite{MG/ME} such as the Standard Model and the
Minimal SUSY Standard Model\cite{MSSM}. There is also a
template for further extension by users, User Mode (usrmod). Using this template, we can add new particles and their
interactions to MadGraph. We make use of this utility and introduce a
model which we call the CFQCD Model.

\begin{figure}
\centering
\resizebox{0.3\textwidth}{!}{%
\includegraphics{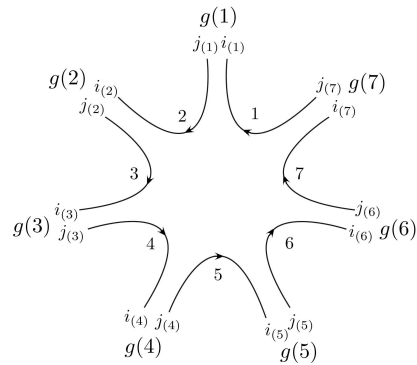}
}
\caption{The color-flow diagram of the process (\ref{eq:gluonprocess})
 with the color flow (\ref{eq:gluoncf}). The number, 1 to 7, counts the
 Kronecker's delta's, or the color-flow lines.}
\label{fig:cfgluon}
\end{figure}

In the CFQCD Model, we introduce gluons and quarks as {\it new}
particles labeled by indices that dictate the color flow, such that the
diagrams for partial amplitudes are generated according to the Feynman rules of
Fig.\,\ref{fig:diagrams}. We need one index for quarks and two indices for gluons, such as $u_k$,
$\overline{u}_k$ and $g_{ji}$. The Abelian gluon, $g_0$, does not
have an index.
 The index labels all possible color flows, and it
 runs from 1 to $m$, where
\begin{equation}
  m = {\rm the\,\, number\,\, of\,\, gluons} + {\rm the\,\, number\,\,
  of\,\, quark\,\, lines}.
\label{eq:m}
\end{equation}
This number $m$ is the number of Kronecker's delta symbols which dictate the color flow.
As an example, let us consider the purely gluonic process
\begin{equation}
g(1)\,g(2)\rightarrow g(3)g(4)g(5)g(6)g(7),
\label{eq:gluonprocess}
\end{equation}
for which we need seven delta's to specify the color flow:
\begin{equation}
(\delta_7)^{i_{(7)}}_{j_{(6)}}(\delta_6)^{i_{(6)}}_{j_{(5)}}
(\delta_5)^{i_{(5)}}_{j_{(4)}}(\delta_4)^{i_{(4)}}_{j_{(3)}}(\delta_3)^{i_{(3)}}_{j_{(2)}}
(\delta_2)^{i_{(2)}}_{j_{(1)}}(\delta_1)^{i_{(1)}}_{j_{(7)}}.
\label{eq:gluoncf}
\end{equation}
Here the numbers in parentheses label gluons
whereas the numbers in the sub-indices of Kronecker's delta's count the color-flow lines, 1 to $m=7$ as depicted in
 Fig.\,\ref{fig:cfgluon}.
 In CFQCD, we label
 gluons according to their flowing-out, {\bf 3}, and flowing-in,
 $\overline{{\bf 3}}$,
 color-flow-line numbers, such that the partial amplitude
 with the color flow (\ref{eq:gluoncf}) is generated as the
 amplitude for the process
\begin{equation}
g_{32}(2)\,g_{21}(1)\rightarrow
 g_{17}(7)\,g_{76}(6)\,g_{65}(5)\,g_{54}(4)\,g_{43}(3).
 \label{eq:process0}
\end{equation}
This is the description of the process (\ref{eq:gluonprocess}) in our
CFQCD Model, and we let MadGraph generate the corresponding partial
amplitudes, such as $A(1,\cdots,n)$ in eq.~(\ref{eq:amplitude}), as the
helicity amplitudes for the process (\ref{eq:process0}). This index number assignment has one-to-one correspondence
with the color flow (\ref{eq:gluoncf}). For instance, $g_{32}(2)$
in the process (\ref{eq:process0}) denotes the contribution of the $U(3)$ gluon
${\cal G}^{\mu}(p_2,\lambda _2)_{j(2)}^{i(2)}$ to the partial amplitude where its
${\bf \bar{3}}$ index $i(2)$ terminates the color-flow line 2, and the
{\bf 3} index $j(2)$ starts the new color-flow line 3. It should also be noted
that we number the color-flow lines in the ascending order along the
color flow, starting from the color-flow line 1 and ending with the
color-flow line $m$. This numbering scheme plays an important role in defining
the interactions among CFQCD gluons and quarks.

\begin{figure}
\resizebox{0.49\textwidth}{!}{%
\includegraphics{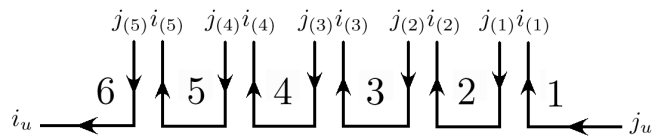}
}
\caption{The color-flow diagram for $u\overline{u}\rightarrow
 5g$ process. The numbers in the diagram shows the color-flow-line number.}
\label{fig:cfdiagram}
\end{figure}

\begin{figure}
\resizebox{0.49\textwidth}{!}{%
\includegraphics{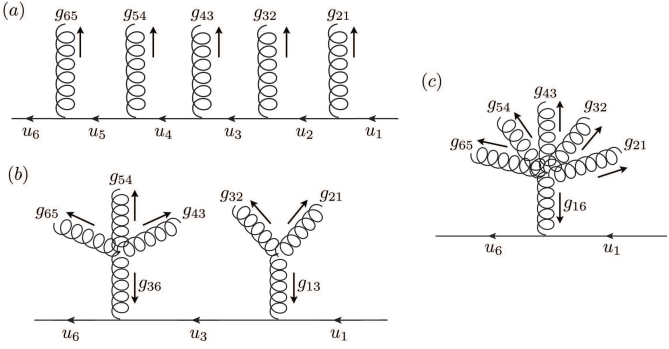}
}
\caption{A few representative Feynman diagrams in the color-flow basis
 for the color flow of Fig.\,\ref{fig:cfdiagram}. Arrows indicate the
 quantum number flows of gluons and quarks.}
\label{fig:repdiagrams}
\end{figure}

Let us now examine the case for one quark line. For the 5-jet production process
\begin{equation}
u\overline{u}\rightarrow g(1)\,g(2)\,g(3)\,g(4)\,g(5),
\label{eq:process1}
\end{equation}
the color-flow index should run from 1 to 6
according to the rule (\ref{eq:m}). Indeed the color flow
\begin{equation}
 (\delta_6)^{i_u}_{j_{(5)}}(\delta_5)^{i_{(5)}}_{j_{(4)}}
(\delta_4)^{i_{(4)}}_{j_{(3)}}
(\delta_3)^{i_{(3)}}_{j_{(2)}}(\delta_2)^{i_{(2)}}_{j_{(1)}}
(\delta_1)^{i_{(1)}}_{j_u}
\label{eq:deltas}
\end{equation}
corresponds to the process
 \begin{equation}
 u_1\,\overline{u}_6\rightarrow
g_{65}(5)\,g_{54}(4)\,g_{43}(3)\,g_{32}(2)\,g_{21}(1).
\label{eq:process2}
\end{equation}
We show in Fig.\,\ref{fig:cfdiagram} the color-flow diagram for this
process. This is just that of the 6-gluon process cut at the
$g_{16}$ gluon, where $g_{16}$ gluon is replaced by the quark pair, $u_1$
and $\overline{u}_6$. Shown in Fig.\,\ref{fig:repdiagrams} are a few representative Feynman
diagrams contributing to the process (\ref{eq:process2}).
Both the external and internal quarks and gluons in the CFQCD model are
shown explicitly along the external and propagator lines. Following the
MadGraph convention, we use flowing-out quantum numbers for gluons while
the quantum numbers are along the fermion number flow for quarks. Gluon
propagators attached to quark lines are named by their color-flow
quantum numbers along arrows. The diagram (a) contains
only $qqg$ vertices, (b) contains an off-shell 4-point gluon current,
{\tt jgluo4}, or a 4-point gluon amplitude, {\tt gluon4}, and (c) contains either an off-shell 6-point gluon current, {\tt
jgluo6}, or a 6-point gluon amplitude, {\tt gluon6}.

\begin{figure}
\resizebox{0.49\textwidth}{!}{%
\includegraphics{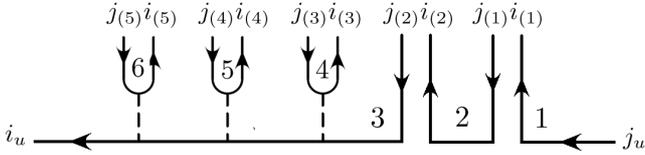}
}
\caption{A color-flow diagram for $u\overline{u}\rightarrow
 5g$ process with Abelian gluons. The numbers in the diagram shows
 the
 color-flow-line number.}
\label{fig:qqbar3g0cf}
\end{figure}

In CFQCD, not only $U(3)$ gluons but also the Abelian gluon
contributes to the processes with quark lines. For the
$u\overline{u}\rightarrow 5g$ process (\ref{eq:process1}), 1 to 5
gluons can be Abelian gluons, $g_0$. If the number of Abelian
gluons is $k$, the color flow reads
\begin{align}
 (\delta_6)_{j_5}^{i_5}\cdots
 (\delta_{6-k+1})_{j_{(5-k+1)}}^{i_{(5-k+1)}}\cdot
 (\delta_{6-k})_{j_{(5-k)}}^{i_u}\cdots (\delta_1)_{j_u}^{i_{(1)}}.
 \label{eq:cfabelian}
\end{align}
When $k=5$, all the five gluons are Abelian, and the first (the right-most)
color flow should be $(\delta_1)_{j_u}^{i_u}$, just as in
$u\overline{u}\rightarrow 5$ photons. In Fig.\,\ref{fig:qqbar3g0cf}, we
show the color-flow diagram for the process (\ref{eq:process1}) with
the color flow (\ref{eq:cfabelian}) for $k=3$ as an
 example.

For the process with two quark lines,
\begin{equation}
 ud\rightarrow ud\,g(1)g(2)g(3),
  \label{eq:ududg3}
\end{equation}
 the color-flow index should run from 1 to 5.
All the possible color flows are obtained from the comb-like diagram for
the one-quark-line process shown in Fig.\,\ref{fig:cfdiagram}, by cutting
one of the gluons into a quark pair, such as $g_{k+1,k}$ to $d_{k+1}$
and $\overline{d}_k$. Then the first color flow starting from $u_1$ ends at $\overline{d}_k$, and
the new color flow starts with $d_{k+1}$ which ends at
$\overline{u}_m$. An example of such color flow for $k=4$ and $m=5$ reads
\begin{equation}
 (\delta_5)^{i_u}_{j_d}\cdot(\delta_4)^{i_d}_{j_{(3)}}(\delta_3)^{i_{(3)}}_{j_{(2)}}
(\delta_2)^{i_{(2)}}_{j_{(1)}}(\delta_1)^{i_{(1)}}_{j_u}.
\end{equation}
The CFQCD model computes the partial amplitude for the above color flow
as the helicity amplitude for the process
\begin{equation}
u_1d_5\rightarrow u_5d_4\,g_{43}(3)\,g_{32}(2)\,g_{21}(1)\,.
\end{equation}
The other possible color-flow processes without Abelian gluons are
\begin{align}
u_1d_4\rightarrow u_5d_3\,g_{32}(2)\,g_{21}(1)\,g_{54}(3),\label{eq:cfudud3g3}\\
u_1d_3\rightarrow u_5d_2\,g_{21}(1)\,g_{54}(3)\,g_{43}(2),\\
u_1d_2\rightarrow u_5d_1\,g_{54}(3)\,g_{43}(2)\,g_{32}(1),
\end{align}
for $k=3,2,1$, respectively. As in the single quark line case, all the
external gluons can also be Abelian gluons, and we should sum over
contributions from external Abelian gluons. For instance, if the
gluon $g(3)$ in the process (\ref{eq:cfudud3g3}) is Abelian, the color flow becomes
\begin{equation}
 (\delta_5)^{i_{(3)}}_{j_{(3)}}\cdot (\delta_4)^{i_u}_{j_d}\cdot
  (\delta_3)^{i_d}_{j_{(2)}}(\delta_2)^{i_{(2)}}_{j_{(1)}}
  (\delta_1)^{i_{(1)}}_{j_u},
\end{equation}
and the corresponding partial amplitude is calculated for the process
\begin{equation}
u_1d_4\rightarrow u_4d_3\,g_{32}(2)\,g_{21}(1)\,g_0(3)
\end{equation}
in CFQCD.

When there are more than one quark line, the Abelian gluon can be
exchanged between two quark lines, and the color flow along each quark
line is disconnected. For instance, the color flow
\begin{equation}
 (\delta_5)^{i_d}_{j_{(3)}}(\delta_4)^{i_{(3)}}_{j_d}\cdot
  (\delta_3)^{i_u}_{j_{(2)}}(\delta_2)^{i_{(2)}}_{j_{(1)}}(\delta_1)^{i_{(1)}}_{j_u}
\end{equation}
is obtained when the Abelian gluon is exchanged between the $u$-quark and
$d$-quark lines. The corresponding partial amplitude is obtained for the
CFQCD process
  \begin{equation}
u_1d_4\rightarrow u_3d_5\,g_{32}(2)\,g_{21}(1)\,g_{54}(3).
\label{eq:g0exchange}
  \end{equation}
Note that in the process (\ref{eq:g0exchange}) each flow starts from and ends
with a quark pair which belongs to the same fermion line.

We have shown so far that we can generate diagrams with definite
color flow by assigning integer labels to quarks, $q_k$, and gluons, $g_{kl}$, such that the labels $k$ and $l$ count the color-flow lines
whose maximum number $m$ is the sum of the number of external gluons and
the numbers of quark lines (quark pairs), see eq.~(\ref{eq:m}).

\subsection{New particles and their interactions in the CFQCD Model}
\label{particle}
In Table~\ref{tb:particles}, we list all the {\it new} particles in the CFQCD
model for $m=6$ in eq.~(\ref{eq:m}).
\begin{table*}
\centering
\begin{tabular}{ccccccccc}
\hline\hline
particle & anti-particle & type & line & mass & width & color & label &
 code\\
\hline\\
$U(3)$ gluons\\
{\tt g13}&{\tt g13}& {\tt V}&{\tt C}& {\tt ZERO} &{\tt ZERO}&{\tt
			 S}&{\tt g13}&{\tt 21}\\
{\tt g14}&{\tt g14}& {\tt V}&{\tt C}& {\tt ZERO} &{\tt ZERO}&{\tt
			 S}&{\tt g14}&{\tt 21}\\
{\tt g15}&{\tt g15}& {\tt V}&{\tt C}& {\tt ZERO} &{\tt ZERO}&{\tt
			 S}&{\tt g15}&{\tt 21}\\
{\tt g16}&{\tt g16}& {\tt V}&{\tt C}& {\tt ZERO} &{\tt ZERO}&{\tt
			 S}&{\tt g16}&{\tt 21}\\
{\tt g21}&{\tt g21}& {\tt V}&{\tt C}& {\tt ZERO} &{\tt ZERO}&{\tt
			 S}&{\tt g21}&{\tt 21}\\
{\tt g24}&{\tt g24}& {\tt V}&{\tt C}& {\tt ZERO} &{\tt ZERO}&{\tt
			 S}&{\tt g24}&{\tt 21}\\
{\tt g25}&{\tt g25}& {\tt V}&{\tt C}& {\tt ZERO} &{\tt ZERO}&{\tt
			 S}&{\tt g25}&{\tt 21}\\
{\tt g26}&{\tt g26}& {\tt V}&{\tt C}& {\tt ZERO} &{\tt ZERO}&{\tt
			 S}&{\tt g26}&{\tt 21}\\
\vdots&\vdots&\vdots&\vdots&\vdots&\vdots&\vdots&\vdots&\\
{\tt g62}&{\tt g62}& {\tt V}&{\tt C}& {\tt ZERO} &{\tt ZERO}&{\tt
			 S}&{\tt g62}&{\tt 21}\\
{\tt g63}&{\tt g63}& {\tt V}&{\tt C}& {\tt ZERO} &{\tt ZERO}&{\tt
			 S}&{\tt g63}&{\tt 21}\\
{\tt g64}&{\tt g64}& {\tt V}&{\tt C}& {\tt ZERO} &{\tt ZERO}&{\tt
			 S}&{\tt g64}&{\tt 21}\\
{\tt g65}&{\tt g65}& {\tt V}&{\tt C}& {\tt ZERO} &{\tt ZERO}&{\tt
			 S}&{\tt g65}&{\tt 21}\\\\
Abelian gluon\\
{\tt g0}&{\tt g0}& {\tt V}&{\tt W}& {\tt ZERO} &{\tt ZERO}&{\tt
			 S}&{\tt g0}&{\tt 21}\\\\
Quarks\\
{\tt u1}&{\tt u1{\footnotesize $\sim$}}& {\tt F}&{\tt S}& {\tt ZERO} &{\tt ZERO}&{\tt
			 S}&{\tt u1}&{\tt 2}\\
{\tt u2}&{\tt u2{\footnotesize $\sim$}}& {\tt F}&{\tt S}& {\tt ZERO} &{\tt ZERO}&{\tt
			 S}&{\tt u2}&{\tt 2}\\
\vdots&\vdots&\vdots&\vdots&\vdots&\vdots&\vdots&\vdots&\\
{\tt u6}&{\tt u6{\footnotesize $\sim$}}& {\tt F}&{\tt S}& {\tt ZERO} &{\tt ZERO}&{\tt
			 S}&{\tt u6}&{\tt 2}\\
\hline\hline
\end{tabular}
\caption{New particles of the CFQCD model when $m =$ the number of gluons
 $+$ the number of quark lines $=6$. The list is shown in the format of {\tt
 particles.dat} in the usrmod of MadGraph\cite{MG/ME}.}
\label{tb:particles}
\end{table*}
 The list is shown in the format of
{\tt particles.dat} in the usrmod of MadGraph\cite{MG/ME}. As
explained above, the $U(3)$ gluons have two color-flow indices,
$g_{kl}$,
while the Abelian gluon, $g_0$, has no
color-flow index. They are vector bosons (type$=${\tt V}), and we use
curly lines (line $=$ {\tt C}) for $U(3)$ gluons while wavy lines (line
$=$ {\tt W}) for the Abelian gluon. The `color' column of the list is
used by MadGraph to perform color summation by using the color-ordered
basis. Since we sum over the color degrees of freedom by summing the {\bf
3} and $\overline{{\bf 3}}$ indices ($j$'s and $i$'s) over all possible color
flows explicitly, we declare all our new particles as singlets (color
$=$ {\tt S})\footnote{If we declare all CFQCD gluons as octet and
quarks as triplets, MadGraph generates color-factor matrices in the
color-ordered basis for each
process, which is not only useless in CFQCD but also consumes memory space.}.
The last column gives the PDG code for the particles, and
all our gluons, including the Abelian gluon, are given the number 21.

All gluons, not only the Abelian gluon but also $U(3)$ gluons, are declared as Majorana particles (particle and anti-particle are the same)
in CFQCD. We adopt this assignment in order to avoid generating
gluon propagators between multi-gluon vertices. Such a propagator is made from a particle coming from one of the vertices and its anti-particle from the other. Since we define the anti-particle of a $U(3)$ gluon as the gluon itself, the color flow of the propagator should be flipped as shown in Fig. \ref{fig:majoprop}, according to the gluon naming scheme explained in the previous subsection.  Therefore,
CFQCD does not give diagrams with gluon propagator in "color-flow
conserving" amplitudes.
\begin{figure}
\centering
\resizebox{0.5\textwidth}{!}{%
\includegraphics{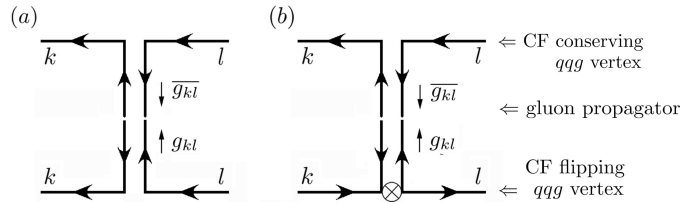}
}
\caption{'Majorana' $U(3)$ gluon propagator in CFQCD. $k$ and $l$ denote the color-flow-line number. Since MadGraph propagator connects a flowing-out particle with a flowing-out anti-particle, it necessarily flips the color flow, as shown along the vertical lines. It either breaks the color-flow conservation, (a), and hence cannot appear inside pure gluon
 amplitudes, or it contributes to amplitudes with quark lines
 with the help of the color-flow flipping $qqg$ vertex, denoted
 by a cross symbol in. (b).}
\end{figure}
For example, to form a $g_{13}$ gluon propagator between two multi-gluon vertices, we need a $g_{13}$ gluon coming out from one of the vertices and a $g_{31}$ gluon from the other, according to the gluon naming scheme explained in the previous subsection. However, because a $g_{31}$ gluon is not an anti-particle of $g_{13}$ but a different particle in our particle difinition, Table~\ref{tb:particles} , $g_{13}$ gluon propagator can not be formed. 
$U(3)$ gluon propagators attached to quark lines
are allowed by introducing the color-flow non-conserving $qqg$ couplings
that effectively recover a color-flow connection. See discussions on $qqg$
vertices at the end of this section.

In Table~\ref{tb:particles}, we list $u$-quarks, $u_k$, and its
anti-particles, $\overline{u}_k$, with the color-flow index $k=1$
to $6$. They are all femions (type $=$ {\tt F}) for which we use solid
lines (line $=$ {\tt S}) in the diagram. Their colors are declared as
singlets (color $=$ {\tt S}) as explained above. The list should be
extended to the other five quarks, such as $d_k$ and $\overline{d}_k$ for down and anti-down quarks.

Before closing the explanation of {\it new} particles in CFQCD, let us note
 that the number of $U(3)$ gluons needed for $m$-gluon processes is $m(m-2)$. This follows from
our color-flow-line numbering scheme, which counts the successive
color-flow lines in the
ascending order along the color flow as depicted in
Fig.\,\ref{fig:cfgluon}. This is necessary to avoid
double counting of the same color flow. According to this rule, the only
gluonic vertex possible for $m=3$ is the one among $g_{13}$, $g_{32}$ and
$g_{21}$. Although $g_{31}$ can appear
 for processes
 with $m\geq4$, $g_{12}$ and $g_{23}$ can
never appear. Generalizing this rule, we find that $g_{kl}$'s with $l=k+1$ mod($m$) as well as $l=k$ cannot appear and hence we need only $m(m-2)$ gluons in CFQCD.

In this paper, we report results up to 5-jet production
processes. Purely gluonic $gg\rightarrow  5g$ process has seven external
gluons, and $m=7$ is necessary only for this process. According to the
above rule, there is only one interaction for this process, which is
the one among $g_{17}$,
$g_{76}$, $g_{65}$, $g_{54}$, $g_{43}$, $g_{32}$ and
$g_{21}$. Therefore, we should
 add $g_{17}$ and $g_{76}$ to Table~\ref{tb:particles} in order to
 compute $gg\rightarrow 5g$ amplitudes.

Next, we list all the interactions among CFQCD particles. Once the
 interactions
 among particles of a
user defined model are given, MadGraph generates Feynman diagrams for an arbitrary
process and the corresponding HELAS amplitude code. In CFQCD, we
introduce $n$-point gluon interactions and let MadGraph
generate a code which calls just one HELAS subroutine (either {\tt gluon\#}
or {\tt jgluo\#} with {\tt \#} $=n$, which makes use of the recursion relations of
eqs.~(\ref{eq:amplitude}) or (\ref{eq:off-shell}), respectively) for each
 $n$-point vertex. In addition, we have
 quark-quark-gluon vertices. We list the interactions in the descending order of the
 number of participating particles in Table \ref{7} to \ref{3}.
\begin{table}
\centering
{\scriptsize
\caption{\normalsize 7-point vertices}
\label{7}
\begin{tabular}{ccc}
\hline
\hline
particles&cpl1 $\cdots$ cpl5&type1 $\cdots$ type5\\
\hline\\
 {\tt g17 g76 g65 g54 g43 g32 g21}&{\tt G2}\;\, $\cdots$\;\, {\tt
     G2}&{\tt QCD}\;\; $\cdots$\;\; {\tt QCD}\\\\
\hline
\hline
\end{tabular}
\vspace{2em}
\caption{\normalsize 6-point vertices}
\label{6}
\begin{tabular}{ccc}
\hline
\hline
particles&cpl1 $\cdots$ cpl4&type1 $\cdots$ type4\\
\hline\\
{\tt g16 g65 g54 g43 g32 g21}&{\tt G2}\;\, $\cdots$\;\, {\tt G2}& {\tt
	 QCD}\;\; $\cdots$\;\; {\tt QCD}\\\\
\hline
\hline
\end{tabular}
\vspace{2em}
\caption{\normalsize 5-point vertices}
\label{5}
\begin{tabular}{ccc}
\hline
\hline
particles&cpl1\, cpl2\, cpl3&type1\,\,type2\,\,type3\\
\hline\\
{\tt g15 g54 g43 g32 g21}&{\tt G2}\hspace{1.2em}{\tt
     G2}\hspace{1.2em}{\tt G2}&{\tt QCD}\hspace{1.2em}{\tt
	 QCD}\hspace{1.2em}{\tt QCD}\\
{\tt g26 g65 g54 g43 g32}&{\tt G2}\hspace{1.2em}{\tt
     G2}\hspace{1.2em}{\tt G2}&{\tt QCD}\hspace{1.2em}{\tt
	 QCD}\hspace{1.2em}{\tt QCD}\\\\
\hline
\hline
\end{tabular}
\vspace{2em}
\caption{\normalsize 4-point vertices}
\label{4}
\begin{tabular}{ccc}
\hline
\hline
particles&cpl1 cpl2&type1 type2\\
\hline\\
{\tt g14 g43 g32 g21}&{\tt G2}\;\;\;\;{\tt G2}&{\tt QCD}\;\;\;\;{\tt
	 QCD}\\
{\tt g25 g54 g43 g32}&{\tt G2}\;\;\;\;{\tt G2}&{\tt QCD}\;\;\;\;{\tt QCD}\\
{\tt g36 g65 g54 g43}&{\tt G2}\;\;\;\;{\tt G2}&{\tt QCD}\;\;\;\;{\tt QCD}\\\\
{\tt g15 g53 g32 g21}&{\tt G2}\;\;\;\;{\tt G2}&{\tt QCD}\;\;\;\;{\tt QCD}\\
{\tt g15 g54 g42 g21}&{\tt G2}\;\;\;\;{\tt G2}&{\tt QCD}\;\;\;\;{\tt QCD}\\
{\tt g15 g54 g43 g31}&{\tt G2}\;\;\;\;{\tt G2}&{\tt QCD}\;\;\;\;{\tt QCD}\\\\

\hline
\hline
\end{tabular}
}
\vspace{1em}
 \begin{flushleft}
 Table~3 to 6: {\tt G2}$=-g_s/\sqrt{2}$ is the coupling of gluon
 vertices and defined as a real, according to the HELAS convention. The list
  is shown in the format of {\tt interactions.dat} in the usrmod of MadGraph\cite{MG/ME}.
  \end{flushleft}
 \label{tb:interactions}
 \end{table}
 \begin{table}
 \centering
\caption{3-point vertices}
\label{3}
\begin{tabular}{ccc}
\hline
\hline
particles&cpl&type\\
\hline\\
$g$-$g$-$g$ vertices&&\\
{\tt g13 g32 g21}&{\tt G2}&{\tt QCD}\\
{\tt g24 g43 g32}&{\tt G2}&{\tt QCD}\\
{\tt g35 g54 g43}&{\tt G2}&{\tt QCD}\\
{\tt g46 g65 g54}&{\tt G2}&{\tt QCD}\\\\
{\tt g14 g42 g21}&{\tt G2}&{\tt QCD}\\
{\tt g14 g43 g31}&{\tt G2}&{\tt QCD}\\\\
{\tt g15 g52 g21}&{\tt G2}&{\tt QCD}\\
{\tt g15 g54 g41}&{\tt G2}&{\tt QCD}\\
{\tt g25 g53 g32}&{\tt G2}&{\tt QCD}\\
{\tt g25 g54 g42}&{\tt G2}&{\tt QCD}\\\\

$q$-$q$-$g$ vertices\\
{\tt u1 u3 g13}&{\tt GG2}&{\tt QCD}\\
{\tt u1 u3 g31}&{\tt GG2}&{\tt QCD}\\
{\tt u3 u1 g31}&{\tt GG2}&{\tt QCD}\\
{\tt u1 u4 g14}&{\tt GG2}&{\tt QCD}\\
{\tt u1 u4 g41}&{\tt GG2}&{\tt QCD}\\
{\tt u4 u1 g41}&{\tt GG2}&{\tt QCD}\\
$\vdots$&&\\
{\tt u6 u7 g76}&{\tt GG2}&{\tt QCD}\\
{\tt u7 u6 g76}&{\tt GG2}&{\tt QCD}\\\\
\hspace{1.5em}$q$-$q$-$g_0$ vertices\\
{\tt u1 u1 g0}&{\tt GG0}&{\tt QCD}\\
{\tt u2 u2 g0}&{\tt GG0}&{\tt QCD}\\
$\vdots$&&\\
{\tt u6 u6 g0}&{\tt GG0}&{\tt QCD}\\\\
\hline
\hline
\end{tabular}
 \vspace{0.5em}
  \begin{flushleft}
{\tt G2}$=-g_s/\sqrt{2}$, {\tt
 GG2}$=(-g_s/\sqrt{2},-g_s/\sqrt{2})$ and {\tt
 GG0}$=(g_s/(\sqrt{2}N),g_s/(\sqrt{2}N))$
 are the couplings of each
 vertex. {\tt G2} is defined as a real and {\tt GG2} and {\tt GG0} are
  defined as two dimensional complex arrays, according to the HELAS convention. The list
  is shown in the format of {\tt interactions.dat} in the usrmod of MadGraph\cite{MG/ME}.
  \end{flushleft}
 \label{tb:interactions}
\end{table}

First, we show the 7-point interaction in Table~\ref{7} in the format of
 {\tt
interactions.dat} of usrmod in MadGraph\cite{MG/ME}. As discussed above,
 it is needed  only for generating $gg\rightarrow 5g$ amplitudes. The vertex is
proportional to $g_s^5$, the fifth power of the strong coupling constant, and
we give the five couplings, cpl1 to cpl5, as
\begin{equation}
 {\tt G2} =-\frac{g_s}{\sqrt{2}},
\end{equation}
according to the Feynman rules of Fig.\,\ref{fig:diagrams}\footnote{The
couplings in the HELAS codes are defined to be $-i$ times the couplings
of the standard Feynman rule.}, whose
types, type1 to type5, are all {\tt QCD}.

The 6-point interactions appear in $gg\rightarrow 4g$ and also in
$q\overline{q}\rightarrow 5g$ process in this study. Again, only one
interaction is possible among the six gluons, $g_{kl}$, with $k=1$ to
$6$ and $l=k-1$, as shown in Table~\ref{6}. The coupling order is
$g_s^4$ and the four couplings, cpl1 to cpl4, are {\tt G2} all with the type
{\tt QCD}.

The 5-point gluon vertices appear in $gg\rightarrow 3g$,
$q\overline{q}\rightarrow 4g$, $q\overline{q}\rightarrow 5g$ and $qq\rightarrow
qq\,3g$ processes. The first two and the fourth processes have $m=5$, for which
the 5-point gluon vertex is unique as shown in the first row of
Table~\ref{5}. The process $q\overline{q}\rightarrow 5g$ gives $m=6$,
and gluons with the sixth color-flow line can contribute to the 5-point gluon
vertex. Because of the ascending color-flow numbering scheme, only one
additional combination appears as shown in the second row of
Table~\ref{5}. These vertices have $g_s^3$ order and we have cpl$k=$ {\tt G2}
and type$k=$ {\tt QCD} for $k=1,2,3$.

The 4-point gluon vertices appear in $gg\rightarrow gg$
($m=4$), $q\overline{q}\rightarrow (m-1)g$ with $m=4$ to $6$, and in
$qq\rightarrow qq+(m-2)g$ with $m=4$ and $5$ in this study. As above, we
have only one 4\,-point gluon vertex for processes with $m=4$, which is shown in
the first row of Table~\ref{4}. For the process $q\overline{q}\rightarrow
4g$ ($m=5$), one additional vertex appears as shown in the second
row. In case of $q\overline{q}\rightarrow 5g$ ($m=6$), the ordering
$3\rightarrow 4\rightarrow 5\rightarrow 6$ also appears, and it is given
in the third row.

So far, we obtain multiple gluon vertices from the color-flow lines of consecutive numbers, corresponding to the color flows
such as
\begin{align}
  y+1&\rightarrow y+2\rightarrow \cdots \rightarrow y+n,\nonumber\\
    &(0\leq y\leq m-n)
\end{align}
 for $n$-point gluon
vertices. When there are two or more quark lines in
the process, we can also have color flows which skips color-flow-line numbers, such as
\begin{align}
 y+1\rightarrow& \cdots \rightarrow y+n\rightarrow \cdots \rightarrow y+n+d,\nonumber\\
 &(0\leq y \leq m-n-d),\nonumber\\
  &(1 \leq d\leq m-n)
\end{align}
for $n$-point gluon vertices, where $d$ counts the number of
skips. Because gluon propagators do not attach to gluon vertices in CFQCD, this skip can appear only when
two or more gluons from the same vertex are connected to quark lines.
 For example, in the fourth row of Table~\ref{4},
the gluon $g_{15}$ couples to the quark line, $u_5\rightarrow u_1$,
and then
the gluon $g_{53}$ couples to the other quark line, $d_3\rightarrow d_5$, in the
process $ud\rightarrow ud\,gg$. Likewise, $g_{42}$ and $g_{31}$ in the
fifth and the sixth row, respectively, couple to the
second quark line, $d\rightarrow d$, of the process.

As for the 3-gluon vertices, ten vertices listed in
Table~\ref{3} appear in our study. The first four vertices with
successive color-flow numbers appear in processes with one quark line,
$q\overline{q}\rightarrow (m-1)g$
with $m=3$ to $6$, and those with two quark lines, $qq\rightarrow
qq\,(m-2)g$ with $m=4$ to 5. The fifth
and the sixth vertices starting with {\tt g14} appears for $qq\rightarrow qq\,(m-2)g$
with $m=4$, for which only one unit of skip ($d=1$) appears, and
also with $m=5$. The
last four vertices appear only in the $m=5$ process with two quark lines, for which
$d=2$ is possible. In fact, the two vertices starting with {\tt g15}
contain {\tt g52} or {\tt g41} with $d=2$ skips in the color-flow
number. This completes all the gluon self-interactions in CFQCD up to
5-jet production processes.

There are two types of $qqg$ vertices in CFQCD: the couplings of $U(3)$
gluons, $g_{kl}$, and those of the Abelian gluon,
$g_0$. All the $qqg$ couplings for the $u$-quark, $u_1$ to $u_6$,  are
listed in Table~\ref{3}. In the HELAS convention, the
fermion-fermion-boson couplings are two dimensional complex arrays,
where the first and the second components are the couplings of the left-
and the right-hand chirality of the flowing-in fermion. According to the
Feynman rules of Fig.\ref{fig:diagrams}, the couplings are
\begin{equation}
{\tt GG2}=\left(-\frac{g_s}{\sqrt{2}},-\frac{g_s}{\sqrt{2}}\right)
 \end{equation}
 for $U(3)$ gluons and
 \begin{equation}
 {\tt GG0}=\left(\frac{g_s}{\sqrt{2}N},\frac{g_s}{\sqrt{2}N}\right)
  \end{equation}
  for the Abelian gluon; Note  the $-1/N$ factor for the Abelian gluon coupling.

$U(3)$ gluons have interactions as
\begin{equation}
 u_l\; u_k\; g_{kl},
  \label{eq:qqg1}
\end{equation}
 where the fermion number flows from $u_l$ to $u_k$ by emitting the out-going $g_{kl}$ gluon. All the diagrams with on-shell $U(3)$ gluons
 attached to a quark line are obtained from the vertices
 (\ref{eq:qqg1}). Likewise, the Abelian gluon couples to quarks as
\begin{equation}
 u_k\; u_k\; g_0.
\end{equation}

In CFQCD, we generate an $U(3)$ gluon propagator between a quark line and
a gluon vertex and between two quark lines by introducing color-flow
flipping vertices
\begin{equation}
 u_k\; u_l\; g_{kl}
  \label{eq:interaction2}
\end{equation}
as we discussed above. Here, the gluon $g_{kl}$ is emitted from the quark line
 $u_k\rightarrow u_l$, such that $U(3)$ gluons can propagate
 between a quark line and a gluonic vertex and also between
 two quark lines when the other side of the $qqg$ vertex is the
 color-flow conserving one, (\ref{eq:qqg1}). In order to exchange gluons between an arbitrary gluonic vertex and a quark line, these color-flow-flipped $qqg$ vertices should exist for all $U(3)$ gluons. However, since all $U(3)$ gluons have the color-flow conserving vertices (\ref{eq:qqg1}) as well, we find double counting of amplitudes where an $U(3)$ gluon is exchanged between two quark lines. This double counting can be avoided simply by discarding color--flow conserving $qqg$ vertices (\ref{eq:qqg1}) for $k<l$ as shown in Table~7. 

Now we exhaust all the interactions needed in
the calculation of partial amplitudes and ready to generate diagrams and evaluate total
amplitudes in the color-flow
basis.

\section{Total amplitudes and the color summation}
\label{total}
In this section, we discuss how we evaluate the total amplitudes,
eqs.~(\ref{eq:Mn}), (\ref{eq:qqng}) and (\ref{eq:2qqng}), with the
CFQCD model and how we perform the color summation of the total
amplitude squared.

\subsection{$gg\rightarrow ng$ processes}
\label{gg}
 We consider pure gluon processes first. As discussed in section~\ref{review}, the total amplitude of a $n$-gluon process is expressed as
eq.~(\ref{eq:Mn}), and it is the sum of several partial
amplitudes. Because color factors, Kronecker's delta's, for partial
amplitudes is either 1 or 0, the total amplitude for a given color
configuration (a color assignment for each external gluons) consists of a subset of all the partial amplitudes in
eq.~(\ref{eq:Mn}).  Therefore, in order to evaluate the total amplitude for a given
color configuration, we should find all possible color flows for the
configuration, compute their partial amplitudes and simply sum
them up.

As an example, let us consider a $n=5$ case,
\begin{equation}
g(1)g(2)\rightarrow g(3)g(4)g(5).
\end{equation}
 The color configuration is expressed by sets of {\bf 3} and
$\overline{\bf 3}$ indices, $j$ and $i$, respectively, for each gluon;
\begin{equation}
(j_1,i_1)_{(1)}, (j_2,i_2)_{(2)}, \cdots, (j_5,i_5)_{(5)}.
\end{equation}
Here the subscripts of each index pair label the gluon $g(k)$, $k=1$ to 5,
with the momentum $p_k$ and the helicity $\lambda_k$. For instance, let us
examine a color
configuration
\begin{equation}
(1,1)_{(1)}, (1,1)_{(2)},(2,1)_{(3)},(3,2)_{(4)},(1,3)_{(5)}.
\label{eq:colorconfig}
\end{equation}
  One of the possible color flows that gives the above color
  configuration is
\begin{equation}
(\delta_5)^{i_{(5)}}_{j_{(4)}}(\delta_4)^{i_{(4)}}_{j_{(3)}}(\delta_3)^{i_{(3)}}_{j_{(2)}}(\delta_2)^{i_{(2)}}_{j_{(1)}}(\delta_1)^{i_{(1)}}_{j_{(5)}},
\label{eq:cf1}
\end{equation}
 whose associated amplitude can be calculated as the amplitude for the
 CFQCD process\\
\begin{equation}
 g_{32}(2)\,g_{21}(1)\rightarrow \,g_{15}(5) \,g_{54}(4) \,g_{43}(3). \\
\end{equation}
MadGraph generates the corresponding HELAS amplitude code, which calls the
5-gluon amplitude subroutine, {\tt gluon5}.

There is also another color flow
\begin{equation}
(\delta_5)^{i_{(2)}}_{j_{(5)}}(\delta_4)^{i_{(5)}}_{j_{(4)}}(\delta_3)^{i_{(4)}}_{j_{(3)}}(\delta_2)^{i_{(3)}}_{j_{(1)}}(\delta_1)^{i_{(1)}}_{j_{(2)}}
\label{eq:cf2}
\end{equation}
for the color configuration (\ref{eq:colorconfig}). The corresponding partial
amplitude can be obtained by one of the $(n-1)!$ permutations of $(n-1)$ gluon momenta and
helicities:
\begin{align}
&\bigl\{\epsilon^{\mu}(5),\epsilon^{\mu}(4),\epsilon^{\mu}(3),\epsilon^{\mu}(2),\epsilon^{\mu}(1)\bigr\}\nonumber\\
\rightarrow& \bigl\{\epsilon^{\mu}(2),\epsilon^{\mu}(5),\epsilon^{\mu}(4),\epsilon^{\mu}(3),\epsilon^{\mu}(1)\bigr\},
\end{align}
where $\epsilon^{\mu}(i)$ denotes the wave function of the external
 gluon $g(i)$.

 When all the partial amplitudes for each color flow are
 evaluated, we sum them up and obtain the color-fixed total
amplitude, eq.~(\ref{eq:Mn}), for this color configuration. Therefore,
for pure gluon process, we generate the HELAS amplitude once for all to evaluate
the color-fixed total amplitude. The total amplitude is then squared and
 the summation over all color
configurations is done by the
Monte-Carlo method.

\subsection{$q\overline{q}\rightarrow ng$ processes}
\label{qqx}
Next, we discuss the processes with one quark line,
$q\overline{q}\rightarrow ng$. The procedure to compute the color-summed amplitude squared is the same, but we should take into account
the Abelian gluon contributions for processes with quarks. As shown in Fig.\,\ref{fig:qqbarcf}, the Abelian gluons appear
as an isolated color index pair which have the same number. Therefore, we can take into account its contribution by regarding the independent gluon index pair as the
Abelian gluon.

As an example, let us consider the process 
\begin{equation}
u\overline{u}\rightarrow g(1)g(2)g(3)g(4)
\end{equation}
 with the color configuration:
\begin{equation}
  (1,1)_u,(1,1)_{(1)},(2,1)_{(2)},(3,2)_{(3)},(1,3)_{(4)},
\label{eq:colorconfig2}
\end{equation}
where the parenthesis with the subscript '$u$' gives the color charges of
the $u$-quark pair; $(j_u, i_u)_u$ denotes the annihilation of the
$u$-quark with the {\bf 3} charge, $j_u$, and the $\overline{u}$-quark
with the $\overline{\bf 3}$ charge, $i_u$.
Although this color configuration is essentially the same as
(\ref{eq:colorconfig}), and hence has the color flows like
(\ref{eq:cf1}) and (\ref{eq:cf2}), we have an additional color flow for this
process:
\begin{equation}
  (\delta_5)^{i_{(1)}}_{j_{(1)}}(\delta_4)^{i_u}_{j_{(4)}}(\delta_3)^{i_{(4)}}_{j_{(3)}}(\delta_2)^{i_{(3)}}_{j_{(2)}}(\delta_1)^{i_{(2)}}_{j_u}\footnote{As
   a convention, we always count the color-flow-lines from the one which starts from
the {\bf 3} index of the quark, $j_q$, for processes with quark lines.}.
\end{equation}
 Here the
index pair of the first gluon, $(1,1)_{(1)}$ in (\ref{eq:colorconfig2}), forms an independent color
flow, $(\delta_5)^{i(1)}_{j(1)}$, which corresponds to the Abelian
gluon, $g_0(1)$. The CFQCD process for this
color flow reads
\begin{center}
  $u_1\overline{u}_4\rightarrow g_{43}(4) \;g_{32}(3) \;g_{21}(2) \;g_0(1)$.
\end{center}
Summing up, there are three color flows
for the color configuration (\ref{eq:colorconfig2}):
\begin{subequations}
\begin{align}
&{\rm (a)}: u\rightarrow 1\rightarrow 2\rightarrow 3\rightarrow 4\rightarrow \overline{u},\label{eq:a}\\
&{\rm (b)}: u\rightarrow 2\rightarrow 3\rightarrow 4\rightarrow 1\rightarrow \overline{u},\label{eq:b}\\
&{\rm (c)}: u\rightarrow 2\rightarrow 3\rightarrow 4\rightarrow
 \overline{u},\; 1\rightarrow 1\label{eq:c}.
\end{align}
\end{subequations}
The partial amplitude $A_b$ for the color flow $(b)$ is obtained from
that of the color flow (a), $A_a$, by a permutation of gluon
wave functions as in the case of $gg\rightarrow 3g$ amplitudes for the
color configuration (\ref{eq:colorconfig}). The partial amplitude for the color flow $(c)$ is calculated with the Feynman rules of
Fig.\,\ref{fig:diagrams}, and the total amplitude ${\cal M}$ is obtained as
\begin{align}
  {\cal M} &= A_a + A_b -\frac{1}{N}\,A_c,
\end{align}
where the $-1/N$ factor comes from the Abelian gluon as shown in eq.~(\ref{eq:qqng}). When more than one gluon have the same color indices, $j_k=i_k$ in
$(j_k,i_k)_{(k)}$, more Abelian gluons can contribute to the amplitude,
and the factor $\left(-1/N\right)^m$ appears for the partial amplitude with $m$ Abelian gluons. In the CFQCD model introduced in section~\ref{implement}, those factors are automatically taken into account in the HELAS code generated by MadGraph.
\subsection{$qq\rightarrow qq\,(n-2)g$ processes}
\label{qq}
\begin{figure}
\resizebox{0.49\textwidth}{!}{%
\includegraphics{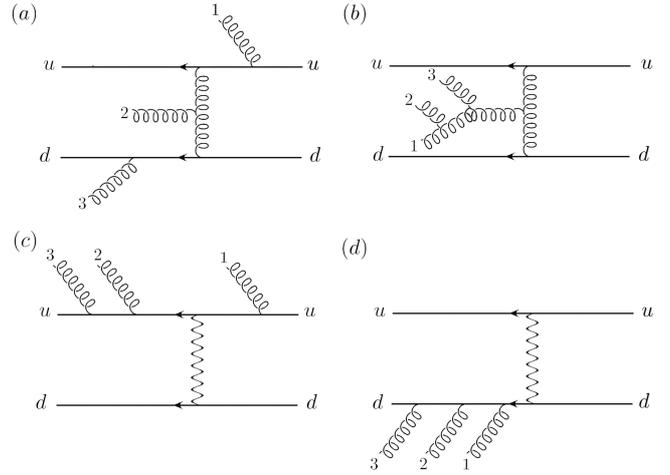}
}
\caption{Feynman diagrams for the color flow (a) to (d). We show one of the
 Feynman diagrams for each color flow.}
\label{fig:abcd}
\end{figure}

Finally, let us discuss the processes with two quark lines. As shown in
Fig.\,\ref{fig:2qqdiagrams}, there are two independent color flows because there are two sets of quark color
indices, $(j_{q_1}, i_{q_1})$ and $(j_{q_2}, i_{q_2})$. Nevertheless, we
can show color configurations and find possible color flows in the same
way as in the $q\overline{q}\rightarrow ng$ cases. Let us consider the process
\begin{equation}
 u d\rightarrow u d \;g(1) g(2) g(3)
\end{equation}
with the color configuration
\begin{equation}
  (1,1)_u,(1,1)_d,(2,1)_{(1)},(3,2)_{(2)},(1,3)_{(3)}
\end{equation}
for illustration. Color charges of $u$-quarks are in the parenthesis
$(j_u,i_u)_u$, and those of $d$-quarks are in $(j_d,i_d)_d$ as in
the $q\overline{q}\rightarrow ng$ processes. All the possible color
flows are
\begin{subequations}
\begin{align}
 {\rm (a)}:\,&(\delta_5)^{i_u}_{j_d}(\delta_4)^{i_d}_{j_{(3)}}(\delta_3)^{i_{(3)}}_{j_{(2)}}(\delta_2)^{i_{(2)}}_{j_{(1)}}(\delta_1)^{i_{(1)}}_{j_u},\\
 &\;(u\rightarrow 1\rightarrow 2\rightarrow 3\rightarrow d, d\rightarrow u)\nonumber\\
 {\rm (b)}:\,&(\delta_5)^{i_u}_{j_{(3)}}(\delta_4)^{i_{(3)}}_{j_{(2)}}(\delta_3)^{i_{(2)}}_{j_{(1)}}(\delta_2)^{i_{(1)}}_{j_d}(\delta_1)^{i_d}_{j_u},\\
 &\;(u\rightarrow d, d\rightarrow 1\rightarrow 2\rightarrow
 3\rightarrow u)\nonumber\\
 {\rm (c)}:\,&(\delta_5)^{i_d}_{j_d}(\delta_4)^{i_u}_{j_{(3)}}(\delta_3)^{i_{(3)}}_{j_{(2)}}(\delta_2)^{i_{(2)}}_{j_{(1)}}(\delta_1)^{i_{(1)}}_{j_u},\\
 &\;(u\rightarrow 1\rightarrow 2\rightarrow 3\rightarrow u, d\rightarrow d)\nonumber\\
 {\rm (d)}:\,&(\delta_5)^{i_d}_{j_{(3)}}(\delta_4)^{i_{(3)}}_{j_{(2)}}(\delta_3)^{i_{(2)}}_{j_{(1)}}(\delta_2)^{i_{(1)}}_{j_d}(\delta_1)^{i_u}_{j_u}.\\
 &\;(u\rightarrow u, d\rightarrow 1\rightarrow 2\rightarrow
 3\rightarrow d)\nonumber
\end{align}
\end{subequations}
We show in Fig.\,\ref{fig:abcd} one representative Feynman diagram for
each color flow, $(a)$ to $(d)$.
The amplitudes for each color flow, $A_a$ to $A_d$, are calculated as
amplitudes for the corresponding CFQCD processes:
\begin{subequations}
\begin{align}
 (a):&\,u_1\,d_5\rightarrow u_5\,d_4 \;g_{43}(3) \;g_{32}(2) \;g_{21}(1),\\
 (b):&\,u_1\,d_2\rightarrow u_5\,d_1 \;g_{54}(3) \;g_{43}(2) \;g_{32}(1),\\
 (c):&\,u_1\,d_5\rightarrow u_4\,d_5 \;g_{43}(3) \;g_{32}(2) \;g_{21}(1),\\
 (d):&\,u_1\,d_2\rightarrow u_1\,d_5 \;g_{54}(3) \;g_{43}(2) \;g_{32}(1).
\end{align}
\end{subequations}
For the CFQCD processes $(c)$ and $(d)$, the color flow lines starting
with the $u$- and $d$-quarks terminate at the same quarks. Such amplitudes are generated
by an exchange of the Abelian gluon, as shown by the
representative diagrams in Fig.\,\ref{fig:abcd}. The total amplitude is
now obtained as
\begin{equation}
 {\cal M}=A_a+A_b-\frac{1}{N}\,A_c-\frac{1}{N}\,A_d,
\end{equation}
according to eq.~(\ref{eq:2qqng}). The color summation of the squared
amplitudes $|{\cal M}|^2$ is performed by the MC method just as in the pure gluon case.

\section{Sample results}
\label{result}
In this section, we present numerical results for several
 multi-jet production processes as a demonstration of our
 CFQCD model on MadGraph. We compute $n$-jet production cross sections from
 $gg\rightarrow ng$, $u\overline{u}\rightarrow ng$ and
 $uu\rightarrow uu+(n-2)g$ subprocesses
 up to $n=5$ in the $pp$ collision at $\sqrt{s}=14$~TeV.
 We define the final state cuts, the QCD coupling constant
 and the parton distribution function exactly as the same
 as those of ref.~\cite{GPU2}, so that we can compare our results
 against those presented in ref.~\cite{GPU2}, which have been
 tested by a few independent calculations.
 Specifically, we select jets (partons) that satisfy
 \begin{subequations}
  \begin{align}
 \label{jet-cuts}
 &|\eta_j| < 2.5,\\
   &p_T(j),\: p_{T_{jk}} > 20~{\mbox{\rm GeV}},
 \end{align}
 \end{subequations}
 where $\eta_j$ and $p_T(j)$ are the pseudo-rapidity and
 the transverse momentum of the parton-$j$,
 and $p_{T_{jk}}$ is the smaller of the relative transverse
 momentum between parton-$j$ and parton-$k$.
 We use CTEQ6L1 parton distribution functions\cite{CTEQ6L1}
 at the factorization scale $Q=20$~GeV and the QCD coupling
 $\alpha_s(Q=20~{\rm GeV})_{\overline{{\rm MS}}}=0.171$.
 Phase space integration and summation over color and helicity
 are performed by an adaptive Monte Carlo (MC) integration
 program BASES\cite{BASES}.

\begin{figure}
\resizebox{0.49\textwidth}{!}{%
\includegraphics{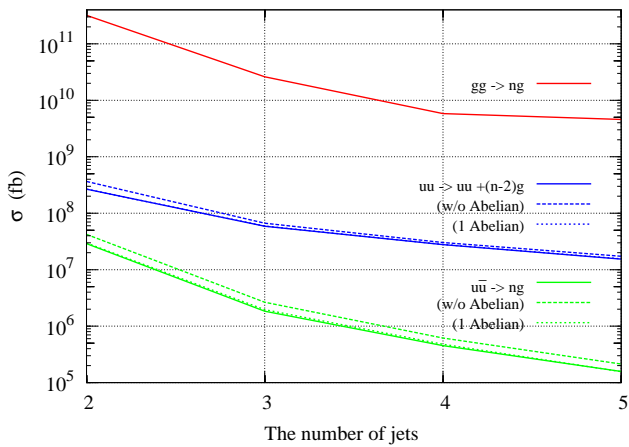}
}
\caption{Total cross sections of $gg\rightarrow ng$ (upper),
 $uu\rightarrow uu+(n-2)g$ (middle) and
 $u\overline{u}\rightarrow ng$ (lower)
 in fb scale for $pp$ collisions at
 $\sqrt{s}=14$ TeV as given in Table~\ref{tb:results}.
 Results shown by dashed lines and dotted lines for $uu$ and $u\overline{u}$
 subprocesses are obtained when Abelian gluon contributions
 are ignored and one Abelian gluon contributions are included, respectively.}
\label{fig:results}
\end{figure}

 Results are shown in Table~\ref{tb:results} and
 Fig.\ref{fig:results}.
\begin{table*}
\small
\begin{center}
\newlength{\mh}
\setlength{\mh}{1em}
\newlength{\mha}
\setlength{\mha}{1.5em}
\begin{tabular}{cccc}
\hline
No. of jets & $gg\rightarrow ng$ & $u\overline{u}\rightarrow ng$ &
 $uu\rightarrow uu\,(n-2)g$ \\
\hline
\rule{0em}{\mh}$2$ & $(3.19\pm 0.00)\times 10^{11}$ & $(2.90\pm 0.00)\times 10^{7}$ &
	     $(2.67\pm 0.01)\times 10^{8}$ \\
 & &$(4.23\pm 0.01)\times 10^{7}$ & $(3.65\pm 0.00)\times 10^{8}$ \\
 & &$(2.97\pm 0.00)\times 10^{7}$ & $(2.67\pm 0.01)\times 10^{8}$ \\
\rule{0em}{\mha}$3$ & $(2.61\pm 0.00)\times 10^{10}$ & $(1.84\pm 0.00)\times 10^{6}$ &
	     $(5.88\pm 0.03)\times 10^{7}$ \\
 & & $(2.65\pm 0.01)\times 10^6$ & $(6.66\pm 0.04)\times 10^7$ \\
 & & $(1.96\pm 0.01)\times 10^6$ & $(5.93\pm 0.03)\times 10^7$ \\
\rule{0em}{\mha}$4$ & $(5.81\pm 0.01)\times 10^9$ & $(4.52\pm0.10)\times 10^5$ & $(2.78\pm
	     0.02)\times 10^7$ \\
 & & $(6.17\pm 0.01)\times 10^5$ & $(3.03\pm 0.03)\times 10^7$ \\
 & & $(4.78\pm 0.01)\times 10^5$ & $(2.80\pm 0.02)\times 10^7$ \\
\rule{0em}{\mha}$5$ &$(4.57\pm 0.01)\times 10^9$ & $(1.57\pm 0.03)\times 10^5$ & $(1.54\pm 0.01)\times 10^7$ \\
 & &$(2.13\pm 0.01)\times 10^5$ & $(1.73\pm 0.01)\times 10^7$ \\
 & &$(1.57\pm 0.04)\times 10^5$ & $(1.56\pm 0.01)\times 10^7$ \\
\hline
\end{tabular}
\vspace{1em}
\caption{Total cross sections of $gg\rightarrow ng$,
 $u\overline{u}\rightarrow ng$ and
 $uu\rightarrow uu+(n-2)g$ ($n\leq5$) in fb scale
 for $pp$ collisions at $\sqrt{s}=14$ TeV,
 when jets satisfy
 $|\eta_j|<2.5$, $p_T(j)>20$ GeV and
 $p_{T_{jk}}>20$ GeV for the smaller of the relative
 transverse momentum between two jets.
 Results in the second row and the third row of each $n$-jet cross section
 are obtained when we ignore Abelian gluon contributions and include one Abelian gluon contributions, respectively.
 Abelian gluons do not contribute to purely gluonic processes.}
\label{tb:results}
\end{center}
\end{table*}

In Table~\ref{tb:results}, the first row for each $n$-jet process gives the
 exact result for the $n$-jet production cross section, while the
 second row shows the cross section when we ignore all the Abelian
 gluon contributions. The third row shows the results where we include up to one Abelian
 gluon contributions: one Abelian gluon emission amplitudes without an
 Abelian gluon exchange for $u\overline{u}\rightarrow ng$ and $uu\rightarrow
 uu+(n-2)g$ processes, and Abelian gluon
 exchange amplitudes without an Abelian gluon emission for $uu\rightarrow
 uu+(n-2)g$ processes.
 All the numerical results for the exact cross
 sections in Table~\ref{tb:results}
 agree with those
 presented in ref.\cite{GPU2} within the accuracy of MC integration.
 In Fig.\,\ref{fig:results}, the multi-jet production cross sections
 are shown for $n=2, 3, 4,$ and 5 jets in units of fb.
 The upper line gives the results for the subprocess $gg\rightarrow ng$.
 The middle lines show those for the subprocess
 $uu\rightarrow uu+(n-2)g$; their amplitudes are obtained from those of
 the
 subprocess
 $ud\rightarrow ud+(n-2)g$ outlined in this study,
 simply by anti-symmetrizing the amplitudes with respect
 to the two external quark wave functions. The solid line gives the
 exact results, while the dashed line gives
 the results when all the Abelian gluon contributions are
 ignored. The dotted line shows the results which include up to one Abelian gluon
 contributions, although it is hard to distinguish from the solid line
 in the figure. Despite their order $1/N_c$ suppression, contributions of Abelian gluons can be significant;
 more than 30\% for $n=2$, about 13\% for $n=3$, while
 about 10\% for $n=$4 and 5.

 The bottom lines show results for the subprocess
 $u\overline{u}\rightarrow ng$. As above, the solid line gives
 the exact cross sections, while
 the dashed line and the dotted line give the results when contributions
 from Abelian gluons are ignored and one Abelian gluon contributions are
 included, respectively.
Unlike the case for $uu \to uu+(n-2)g$ subprocesses,
 the Abelian gluon contributions remain at 30\% level
 even for $n=5$.

  Before closing this section, we would like to give two
 technical remarks on our implementation of CFQCD on MadGraph.
 First, since the present MadGraph\cite{MG/ME} does not allow vertices
 among more than 4 particles, we add HELAS codes for 5, 6,
 and 7 gluon vertices by hand to complete the MadGraph
 generated codes.
 This restriction will disappear once the new version of
 MadGraph, MG5\footnote{The beta version of MadGraph version 5 is
 available from https://launchpad.net/madgraph5.},
 is available, since MG5 accepts
 vertices with arbitrary number of particles.
 Second, we do not expect difficulty in running CFQCD codes
 on GPU, since all the codes we developed (see Appendix)
 follow the standard HELAS subroutine rules.

\section{Conclusions}
\label{conclusion}
 In this paper, we have implemented off-shell recursive
 formulae for gluon currents in the color-flow basis in MadGraph and have
 shown that it is possible to generate QCD amplitudes
 in the color-flow basis by introducing a new model,
 CFQCD, in which quarks and gluons are labeled by
 color-flow numbers.
 We have

 \noindent
 - introduced new subroutines for off-shell recursive
 formulae for gluon currents, the contact 3- and 4-point gluon vertices
 in the color flow basis and the off-shell Abelian gluon current,\\
 - defined new MadGraph model: the CFQCD Model,\\
 - generated HELAS amplitudes for given color flows
 and calculated the color-summed total amplitude squared,
 and \\
 - showed the numerical results for $n$-jet production
 cross sections ($n\leq 5$).

 Although we have studied only up to 5-jet production processes in this paper, it is straightforward to extend
 the method to higher $n$-jet production processes.

\vspace{3em}
\begin{center}
{\bf Acknowledgment}
\end{center}
We would like to thank Fabio Maltoni for his invaluable comments.  This work is supported in part by
 the Grant-in-Aid for Scientific Research (No. 20340064)
 from the Japan Society for the Promotion of Science.\\

\vspace{3em}
\noindent{\bf Appendix: Sample codes for off-shell currents and
amplitudes}\\

In this Appendix, we list HELAS codes for the contact 4-point
 gluon vertex subroutines,
 {\tt ggggcf} and {\tt jgggcf},
 which sums over contributions with definite color-flow.
 In addition, we list HELAS codes for the 5-gluon
 amplitude subroutine, {\tt gluon5}, and
 the 5-point off-shell gluon current subroutine,
 {\tt jgluo5}, as examples of the recursive multi-gluon
 vertices introduced in section~\ref{implement}.
 
 \vspace{2em}
\begin{center}
 {\bf A1.}~{\tt ggggcf}\\
{\tt 
\begin{supertabular}{l}
************************************************\\
     \hspace{0.5em} subroutine ggggcf(g1,g2,g3,g4,g, vertex)\\
\\
      \hspace{0.5em} implicit none\\
\\
      \hspace{0.5em} complex*16    g1(6),g2(6),g3(6),g4(6),vertex\\
      \hspace{0.5em} complex*16 dv1(0:3),dv2(0:3),dv3(0:3),dv4(0:3)\\
      \hspace{0.5em} complex*16 dvertx,v12,v13,v14,v23,v24,v34\\
      \hspace{0.5em} real*8 g\\
\\
      \hspace{0.5em} dv1(0)=dcmplx(g1(1))\\
      \hspace{0.5em} dv1(1)=dcmplx(g1(2))\\
      \hspace{0.5em} dv1(2)=dcmplx(g1(3))\\
      \hspace{0.5em} dv1(3)=dcmplx(g1(4))\\
      \hspace{0.5em} dv2(0)=dcmplx(g2(1))\\
      \hspace{0.5em} dv2(1)=dcmplx(g2(2))\\
      \hspace{0.5em} dv2(2)=dcmplx(g2(3))\\
      \hspace{0.5em} dv2(3)=dcmplx(g2(4))\\
      \hspace{0.5em} dv3(0)=dcmplx(g3(1))\\
      \hspace{0.5em} dv3(1)=dcmplx(g3(2))\\
      \hspace{0.5em} dv3(2)=dcmplx(g3(3))\\
      \hspace{0.5em} dv3(3)=dcmplx(g3(4))\\
      \hspace{0.5em} dv4(0)=dcmplx(g4(1))\\
      \hspace{0.5em} dv4(1)=dcmplx(g4(2))\\
      \hspace{0.5em} dv4(2)=dcmplx(g4(3))\\
      \hspace{0.5em} dv4(3)=dcmplx(g4(4))\\
\\
      \hspace{0.5em} v12= dv1(0)*dv2(0)-dv1(1)*dv2(1)-dv1(2)*dv2(2)\\
      \hspace{0.5em}\& \hspace{1.5em}-dv1(3)*dv2(3)\\
      \hspace{0.5em} v13= dv1(0)*dv3(0)-dv1(1)*dv3(1)-dv1(2)*dv3(2)\\
       \hspace{0.5em}\& \hspace{1.5em}-dv1(3)*dv3(3)\\
      \hspace{0.5em} v14= dv1(0)*dv4(0)-dv1(1)*dv4(1)-dv1(2)*dv4(2)\\
       \hspace{0.5em}\& \hspace{1.5em}-dv1(3)*dv4(3)\\
      \hspace{0.5em} v23= dv2(0)*dv3(0)-dv2(1)*dv3(1)-dv2(2)*dv3(2)\\
       \hspace{0.5em}\& \hspace{1.5em}-dv2(3)*dv3(3)\\
      \hspace{0.5em} v24= dv2(0)*dv4(0)-dv2(1)*dv4(1)-dv2(2)*dv4(2)\\
       \hspace{0.5em}\& \hspace{1.5em}-dv2(3)*dv4(3)\\
      \hspace{0.5em} v34= dv3(0)*dv4(0)-dv3(1)*dv4(1)-dv3(2)*dv4(2)\\
       \hspace{0.5em}\& \hspace{1.5em}-dv3(3)*dv4(3)\\
\\
      \hspace{0.5em} dvertx =(-v14*v23+2.d0*v13*v24-v12*v34)\\
\\
      \hspace{0.5em} vertex = dcmplx(dvertx)*(g*g)\\
\\
      \hspace{0.5em} return\\
      \hspace{0.5em} end\\
************************************************\\
\end{supertabular}
}\\
\vspace{1em}
{\bf A2.}~{\tt jgggcf}\\
{\tt
\begin{supertabular}{l}
************************************************\\
       \hspace{0.5em} subroutine jgggcf(w1,w2,w3,g, jggg)\\
\\
       \hspace{0.5em} implicit none\\
\\
      \hspace{0.5em} double complex w1(6),w2(6),w3(6),jggg(6)\\
      \hspace{0.5em} double complex dw1(0:3),dw2(0:3),dw3(0:3)\\
      \hspace{0.5em} double complex jj(0:3),dv,w32,w13,w21,dg2\\
      \hspace{0.5em} double precision q(0:3,)q2,g\\
\\
      \hspace{0.5em} double precision rOne\\
      \hspace{0.5em} parameter( rOne = 1.0d0 )\\
\\
      \hspace{0.5em} jggg(5) = w1(5)+w2(5)+w3(5)\\
      \hspace{0.5em} jggg(6) = w1(6)+w2(6)+w3(6)\\
\\
      \hspace{0.5em} dw1(0) = dcmplx(w1(1))\\
      \hspace{0.5em} dw1(1) = dcmplx(w1(2))\\
      \hspace{0.5em} dw1(2) = dcmplx(w1(3))\\
      \hspace{0.5em} dw1(3) = dcmplx(w1(4))\\
      \hspace{0.5em} dw2(0) = dcmplx(w2(1))\\
      \hspace{0.5em} dw2(1) = dcmplx(w2(2))\\
      \hspace{0.5em} dw2(2) = dcmplx(w2(3))\\
      \hspace{0.5em} dw2(3) = dcmplx(w2(4))\\
      \hspace{0.5em} dw3(0) = dcmplx(w3(1))\\
      \hspace{0.5em} dw3(1) = dcmplx(w3(2))\\
      \hspace{0.5em} dw3(2) = dcmplx(w3(3))\\
      \hspace{0.5em} dw3(3) = dcmplx(w3(4))\\
      \hspace{0.5em} q(0) = -dble(jggg(5))\\
      \hspace{0.5em} q(1) = -dble(jggg(6))\\
      \hspace{0.5em} q(2) = -dimag(jggg(6))\\
      \hspace{0.5em} q(3) = -dimag(jggg(5))\\
\\
      \hspace{0.5em} q2 = q(0)**2-(q(1)**2+q(2)**2+q(3)**2)\\
\\
      \hspace{0.5em} dg2 = g*g\\

      \hspace{0.5em} dv = rOne/dcmplx(q2)\\
\\
      \hspace{0.5em} w32 = dw3(0)*dw2(0)-dw3(1)*dw2(1)\\
       \hspace{0.5em}\& \hspace{2em}-dw3(2)*dw2(2)-dw3(3)*dw2(3)\\
      \hspace{0.5em} w13 = dw1(0)*dw3(0)-dw1(1)*dw3(1)\\
       \hspace{0.5em}\& \hspace{2em}-dw1(2)*dw3(2)-dw1(3)*dw3(3)\\
      \hspace{0.5em} w21 = dw2(0)*dw1(0)-dw2(1)*dw1(1)\\
       \hspace{0.5em}\& \hspace{2em}-dw2(2)*dw1(2)-dw2(3)*dw1(3)\\
\\
      \hspace{0.5em} jj(0) = dg2*(-dw1(0)*w32+2d0*dw2(0)*w13\\
       \hspace{0.5em}\& \hspace{3em}-dw3(0)*w21)\\
      \hspace{0.5em} jj(1) = dg2*(-dw1(1)*w32+2d0*dw2(1)*w13\\
       \hspace{0.5em}\& \hspace{3em}-dw3(1)*w21)\\
      \hspace{0.5em} jj(2) = dg2*(-dw1(2)*w32+2d0*dw2(2)*w13\\
       \hspace{0.5em}\& \hspace{3em}-dw3(2)*w21)\\
      \hspace{0.5em} jj(3) = dg2*(-dw1(3)*w32+2d0*dw2(3)*w13\\
       \hspace{0.5em}\& \hspace{3em}-dw3(3)*w21)\\
\\
      \hspace{0.5em} jggg(1) = dcmplx(jj(0)*dv)\\
      \hspace{0.5em} jggg(2) = dcmplx(jj(1)*dv)\\
      \hspace{0.5em} jggg(3) = dcmplx(jj(2)*dv)\\
      \hspace{0.5em} jggg(4) = dcmplx(jj(3)*dv)\\
\\
      \hspace{0.5em} return\\
      \hspace{0.5em} end\\
************************************************\\
\end{supertabular}
}

\vspace{1em}
{\bf A3.}~{\tt gluon5}\\
{\tt
\begin{supertabular}{l}
************************************************\\
      \hspace{0.5em} subroutine gluon5 (w1,w2,w3,w4,w5,g,vertex)\\
\\
     \hspace{0.5em} implicit none\\
\\
     \hspace{0.5em} integer i\\
     \hspace{0.5em} complex*16 w(6,12,12),w1(6),w2(6),w3(6),w4(6)\\
     \hspace{0.5em} complex*16 w5(6),z(55),wx(6,55),vertex    \\
     \hspace{0.5em} real*8 g    \\
      \\
     \hspace{0.5em} vertex=(0d0,0d0)     \\
\\
     \hspace{0.5em} do i=1,6\\
     \hspace{1em}    w(i,1,1) = w1(i)\\
     \hspace{1em}    w(i,2,2) = w2(i)\\
     \hspace{1em}    w(i,3,3) = w3(i)\\
     \hspace{1em}    w(i,4,4) = w4(i)\\
     \hspace{1em}    w(i,5,5) = w5(i)\\
     \hspace{0.5em} enddo\\
\\
     \hspace{0.5em} do i=1,3\\
     \hspace{1em} call jggxxx(w(1,i,i),w(1,i+1,i+1),g,\\
     \&\hspace{7em}w(1,i,i+1))     \\
     \hspace{0.5em} enddo\\
\\
      \hspace{0.5em} do i=1,2  \\
      \hspace{1em}    call jggxxx(w(1,i,i+1),w(1,i+2,i+2),g,\\
      \&\hspace{7em}wx(1,1))\\
      \hspace{1em}    call jggxxx(w(1,i,i),w(1,i+1,i+2),g,wx(1,2))    \\
      \hspace{1em}    call jgggcf(w(1,i,i),w(1,i+1,i+1),\\
      \hspace{0.2em}\&\hspace{6.8em}w(1,i+2,i+2),g,wx(1,3)) \\
      \hspace{1em}    call sumw(wx,3,w(1,i,i+2))\\
      \hspace{0.5em} enddo   \\
      \\
      \hspace{0.5em} call gggxxx(w(1,1,3),w(1,4,4),w(1,5,5),g,z(1))   \\
      \hspace{0.5em} call gggxxx(w(1,1,2),w(1,3,4),w(1,5,5),g,z(2))     \\
      \hspace{0.5em} call gggxxx(w(1,1,1),w(1,2,4),w(1,5,5),g,z(3))      \\
      \hspace{0.5em} call ggggcf(w(1,1,2),w(1,3,3),w(1,4,4),\\
      \&\hspace{6.6em}w(1,5,5),g,z(4))\\
      \hspace{0.5em} call ggggcf(w(1,1,1),w(1,2,3),w(1,4,4),\\
      \&\hspace{6.6em}w(1,5,5),g,z(5))  \\
      \hspace{0.5em} call ggggcf(w(1,1,1),w(1,2,2),w(1,3,4),\\
      \&\hspace{6.6em}w(1,5,5),g,z(6))    \\
      \hspace{0.5em} do i=1,6\\
      \hspace{1em}    vertex = vertex+z(i)\\
      \hspace{0.5em} enddo   \\
       \\
      \hspace{0.5em} return \\
      \hspace{0.5em} end\\
************************************************\\
\end{supertabular}
}

\vspace{1em}
{\bf A4.}~{\tt jgluo5}\\
{\tt
\begin{supertabular}{l}
************************************************\\
     \hspace{0.5em} subroutine jgluo5(w1,w2,w3,w4,g, jgluon5) \\
\\
     \hspace{0.5em} implicit none\\
\\
     \hspace{0.5em} integer i\\
     \hspace{0.5em} complex*16 w(6,12,12),w1(6),w2(6),w3(6)\\
     \hspace{0.5em} complex*16 w4(6),wx(6,55),jgluon5(6)   \\
     \hspace{0.5em} real*8 g\\
\\
     \hspace{0.5em} do i=1,6\\
     \hspace{1em}    jgluon5(i) = (0d0,0d0)\\
     \hspace{1em}    w(i,1,1) = w1(i)\\
     \hspace{1em}    w(i,2,2) = w2(i)\\
     \hspace{1em}    w(i,3,3) = w3(i)\\
     \hspace{1em}    w(i,4,4) = w4(i)\\
     \hspace{0.5em} enddo\\
     \\
      \hspace{0.5em} do i=1,3\\
      \hspace{1em}   call jggxxx(w(1,i,i),w(1,i+1,i+1),g,\\
      \&\hspace{7em}w(1,i,i+1))\\
      \hspace{0.5em} enddo\\
\\
       \hspace{0.5em} do i=1,2 \\
       \hspace{1em}   call jggxxx(w(1,i,i+1),w(1,i+2,i+2),g,\\
       \&\hspace{7em}wx(1,1))  \\
       \hspace{1em}    call jggxxx(w(1,i,i),w(1,i+1,i+2),g,wx(1,2))      \\
       \hspace{1em}    call jgggcf(w(1,i,i),w(1,i+1,i+1),\\
       \&\hspace{7em}w(1,i+2,i+2),g,wx(1,3))\\
       \hspace{1em}    call sumw(wx,3,w(1,i,i+2))\\
       \hspace{0.5em} enddo\\
      \\
      \hspace{0.5em} call jggxxx(w(1,1,3),w(1,4,4),g,wx(1,1))  \\
      \hspace{0.5em} call jggxxx(w(1,1,2),w(1,3,4),g,wx(1,2))    \\
      \hspace{0.5em} call jggxxx(w(1,1,1),w(1,2,4),g,wx(1,3))      \\
      \hspace{0.5em} call jgggcf(w(1,1,2),w(1,3,3),w(1,4,4),g,\\
      \&\hspace{6.5em}wx(1,4)) \\
      \hspace{0.5em} call jgggcf(w(1,1,1),w(1,2,3),w(1,4,4),g,\\
      \&\hspace{6.5em}wx(1,5))   \\
      \hspace{0.5em} call jgggcf(w(1,1,1),w(1,2,2),w(1,3,4),g,\\
      \&\hspace{6.5em}wx(1,6))     \\
      \hspace{0.5em} call sumw(wx,6,jgluon5)\\
 \\
     \hspace{0.5em} return\\
     \hspace{0.5em} end\\
************************************************\\
\end{supertabular}
}
\end{center}

\end{document}